\documentclass[10pt,letterpaper]{article}
\usepackage[top=0.85in,left=2.75in,footskip=0.75in]{geometry}

\usepackage{amsmath,amssymb}

\usepackage{changepage}

\usepackage[utf8x]{inputenc}

\usepackage{textcomp,marvosym}

\usepackage{cite}

\usepackage{nameref,hyperref}

\usepackage[right]{lineno}

\usepackage{microtype}
\DisableLigatures[f]{encoding = *, family = * }

\usepackage[table]{xcolor}

\usepackage{array}

\newcolumntype{+}{!{\vrule width 2pt}}

\newlength\savedwidth

\raggedright
\usepackage[aboveskip=1pt,labelfont=bf,labelsep=period,justification=raggedright,singlelinecheck=off]{caption}

\makeatletter
\renewcommand{\@biblabel}[1]{\quad#1.}
\makeatother

\usepackage{lastpage,fancyhdr,graphicx}
\usepackage{epstopdf}
\usepackage{longtable}
\fancyheadoffset[L]{2.25in}
\fancyfootoffset[L]{2.25in}
\begin{document}
\vspace*{0.2in}

\begin{flushleft}
{\Large
\textbf\newline{Bow-tie structure and community identification of global supply chain network} % Please use "sentence case" for title and headings (capitalize only the first word in a title (or heading), the first word in a subtitle (or subheading), and any proper nouns).
}
\newline
\\
Abhijit Chakraborty\textsuperscript{1,*},
Yuichi Ikeda\textsuperscript{2, \Yinyang}
\\
\bigskip
\textbf{1} Graduate School of Simulation Studies, The University of Hyogo, Kobe 650-0047, Japan
\\
\textbf{2} Graduate School of Advanced Integrated Studies in Human Survivability, Kyoto University, Kyoto 606-8306, Japan\\
\bigskip

% Insert additional author notes using the symbols described below. Insert symbol callouts after author names as necessary.
% 
% Remove or comment out the author notes below if they aren't used.
%
% Primary Equal Contribution Note
%\Yinyang These authors contributed equally to this work.

% Additional Equal Contribution Note
% Also use this double-dagger symbol for special authorship notes, such as senior authorship.
%\ddag These authors also contributed equally to this work.

%\textcurrency Current Address: Dept/Program/Center, Institution Name, City, State, Country % change symbol to "\textcurrency a" if more than one current address note
% \textcurrency b Insert second current address 
% \textcurrency c Insert third current address


% Group/Consortium Author Note
%\textpilcrow Membership list can be found in the Acknowledgments section.

% Use the asterisk to denote corresponding authorship and provide email address in note below.
* abhiphyiitg@gmail.com
\Yinyang ikeda.yuichi.2w@kyoto-u.ac.jp

\end{flushleft}
\section*{Abstract}
We study on topological properties of global supply chain network in terms of degree distribution, hierarchical structure, and degree-degree correlation in the global supply chain network. The global supply chain data is constructed by collecting various company data from the web site of Standard \& Poor's Capital IQ platform in 2018.
The in- and out-degree distributions are characterized by a power law of the form $P(k) \propto k^{-\gamma}$ with $\gamma_{in} = 2.42$ and $\gamma_{out} = 2.11$. The clustering coefficient decays $\langle C(k) \rangle \sim k^{-\beta_k}$ with an exponent $\beta_k = 0.46$. The nodal degree-degree correlation $\langle k_{nn}(k) \rangle$ indicates the absence of assortativity. The Bow-tie structure of GWCC reveals that the OUT component is the largest and it consists $41.1\%$ of total firms. The GSCC component comprises $16.4\%$ of total firms. We observe that the firms in the upstream or downstream sides are mostly located a few steps away from the GSCC. 
Furthermore, we uncover the community structure of the network and characterize them according to their location and industry classification. We observe that the largest community consists of consumer discretionary sector mainly based in the US. These firms belong to the OUT component in the bow-tie structure of the global supply chain network.
Finally, we confirm the validity for propositions S1 (short path length), S2 (power-law degree distribution), S3 (high clustering coefficient), S4 (“fit-gets-richer” growth mechanism), S5 (truncation of power-law degree distribution), and S7 (community structure with overlapping boundaries) in the global supply chain network.

%\linenumbers

\section*{Introduction}
National economies are linked by international trade and consequently economic globalization forms a giant economic complex network with strong links, i.e., interactions due to increasing trade. 
Especially if we view the globalized world economy with high resolution or microscopic view, we might notice that the giant economic network is a global supply chain consisting of a huge number of firms. 
On the other hand, it has been known that various collective motions exist in natural and social phenomena. The collective motions are due to strong interactions between constituent elements. Thus, it is expected that various collective motions will emerge in the globalized world economy under trade liberalization.

In the study of supply chain network, a several review papers have been published. 
M.~A.~Bellamy~{\em et. al.}~\cite{Bellamy2012} categorized the study into three themes: network structure (i.e., system architecture), network dynamics (i.e., system behavior), and network strategy (system policy and control). They listed important factors to characterize supply chain network. For instance, factors of network structure are node-level property, network-level property, and link-level property. The factors of network dynamics are stimuli, phenomenon, and sustainability. The factors of network strategy are scope, intent, and governance. 
S.~Perera {\em et. al.}~\cite{Perera2017} surveyed the methodologies to model topology and robustness. They pointed out the limitation of preferential attachment growth model to explain characteristics of the supply chain network and stressed the importance of fitness based growth models \cite{Bell2017} to explain the observed topological characteristics.  
Notable phenomena on the supply chain networks are not only resilience against random failure and targeted attack but also collective motion such as cascading failure or chain bankruptcy.
Y.~Fujiwara studied the chain bankruptcy by analyzing supply chain and bankruptcy data, and Y.~Ikeda developed a agent-based model and ran realistic simulation of the chain bankruptcy caused by a failure of a single firm~\cite{Econophysics2010}.
K.~J.~Mizgier~{\em et. al.}~\cite{Mizgier2012} studied the dynamics of default process in supply chain network using a agent-based model. Based on the simulation, they discussed implication in risk management and policy making.  
L.~Tang~ {\em et. al.}~\cite{Tang2016} developed a theoretical cascading failure model considering  interdependence of firms in supply chain network. They observed a sudden collapse of the interdependence of supply chain network.  
T. Mizuno {\em et. al.}~\cite{mizuno2016structure} have analyzed a  large set of global supply chain data. They have investigated three different types of networks: a customer-supplier network, a licensee-licensor network and a strategic alliance network. The degree distributions of all these three networks show scale free properties characterized by a power law tail. They also observed that all three network shows average path length around six. They have further studied the community structure of undirected versions of the networks using modularity maximization technique~\cite{clauset2004finding}. 

In addition to these studies, E.~J.S.~Hearnshaw~{\em et. al.}~\cite{Hearnshaw2013} have studied the supply chain network in terms of complex network approach and have proposed the following nine propositions:
\begin{itemize}
  \setlength{\parskip}{0cm} 
  \setlength{\itemsep}{0cm} 
  \item S1: \textit{Efficient supply chain systems show a short characteristic path length}
  \item S2: \textit{The nodal degree distribution of efficient supply chain systems follows a power law as indicated by the presence of hub firms}
  \item S3: \textit{Efficient supply chain systems demonstrate a high clustering coefficient}
  \item S4: \textit{The growth of efficient supply chain systems follows “fit-gets-richer" mechanism}
  \item S5: \textit{The power law degree distribution of efficient supply chain systems is truncated}
  \item S6: \textit{The link weight distribution of efficient supply chain systems follows a power law}
  \item S7: \textit{Efficient supply chain systems demonstrate a pronounced community structure with overlapping boundaries}
  \item S8: \textit{The fitness of hub firms determines the resilience of supply chain systems against both random disturbances and targeted attacks}
  \item S9: \textit{Resilient supply chain systems demonstrate a power law distribution for link-weights}
\end{itemize}
The nine propositions are related to path length, power-law degree distribution, clustering coefficient, preferential attachment growth mechanism, truncated power-law connectivity distribution, power-law distribution of node strength, community structure with overlapping boundaries, resilience against random failure and targeted attack, core-periphery structure, respectively. They tried to explain various functions of the supply chain by the structural characteristics of supply chain network.

In order to understand the globalized world economy and to make effective policy recommendations,
it is indispensable to study global supply chain, international trade, business cycle, and economic growth by analyzing global big data using network scientific methodology. 
In this paper, we focus on topological properties of the global supply chain network. 
The study on topological properties of the global supply chain network is the first step to understand the globalized world economy with a  microscopic view.
We study a degree distribution, hierarchical structure, and the degree-degree correlation in the global supply chain network.
We uncover the community structure of the network using map equation method and characterized them according to their location and industry classification. 
Furthermore, the composition of communities in terms of the bow-tie components is analyzed.
Finally, we investigate the validity of the nine propositions on the supply chain network  \cite{Hearnshaw2013} based on the obtained results on the topological properties of global supply chain network.

Our paper is organized as follows. 
In section Data, we briefly describe the global supply chain network data used in this study. The data was collected from Standard \& Poor's Capital IQ platform in 2018. In section Methods, methodologies for the identification of bow-tie structure, the community detection, and the over-expression of bow-tie components are explained. In section Results, the obtained results from the analysis of the global supply chain network data: basic structural properties, bow-tie structure, community structure, and over-expression of bow-tie components are explained using figures and tables. Finally, we investigate the validity of the nine propositions based on the obtained results on the topological properties of global supply chain network. To close, this paper concludes in section Conclusions.

\section*{Data}

The global supply chain data was constructed by collecting various firm data from the web site of Standard \& Poor's (S\&P) Capital IQ platform in 2018.
The data include firm ID, firm name, country and location of firm, primary industry, and sector  as node information. The industrial classification is based on the Global Industry Classification Standard (GICS) which is developed by Morgan Stanley Capital International and S\&P. We have $206$ countries as the location of firms $11$ different sectors of firms,  $158$ primary industries as listed in Table~S1-S3 of Appendix S1.  

The data also include types of business relationship between supplier and customer as link information.
Although the various types of business relationships that come under suppliers are supplier, creditor, franchisor, licensor, landlord, lessor, auditor, transfer agent, investor relations firm, and vendor, the majority of the relationship types are supplier and creditor.
Here the supplier indicates a firm providing the products or services and the creditor indicates a private, public or institutional entity which makes funds available to others to borrow.

In Table~\ref{tab:Allfirms_link}, types of business relationship for all firms are summarized.
We note that the links in the data set are dominated by the business relationship of supply chain.
In Table~\ref{tab:Allfirms}, types of supplier for all firms are summarized.
We note that the suppliers are dominated by private firms and public firms. 
Therefore, the entire characteristics of the data set is reflecting the nature of the global supply chain network.  

\begin{figure}[!h]
\centering
\includegraphics[width=0.7\linewidth]{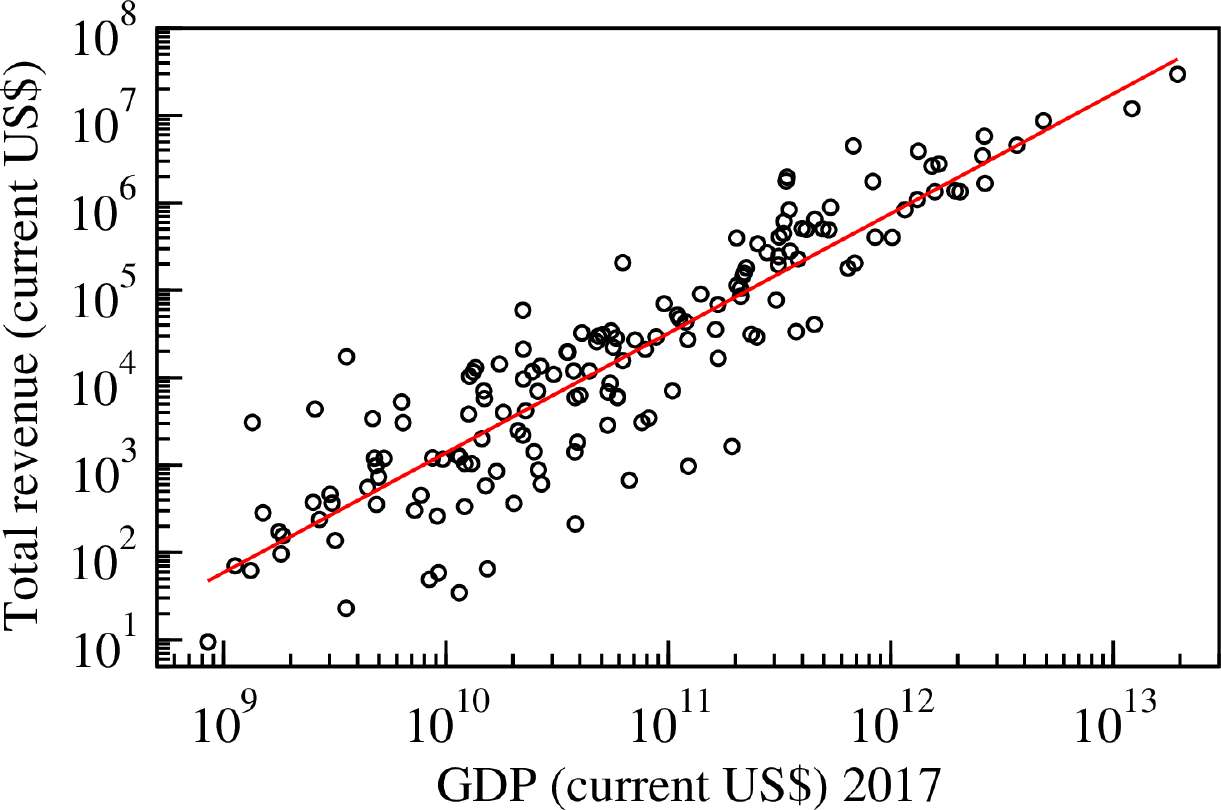}
\caption{{\bf Total revenue generated by the firms within a country is plotted with Country's GDP in current US \$\ for the year $2017$.}
}
\label{fig:GDP_revenue}
\end{figure}

The total number of firms and directed links are $437,453$ and $948,247$, respectively.
Number of firms, total revenue of firms for each country is listed in Table S1 of Appendix S1. Firm distribution for different sectors are listed in Table S2 of Appendix S1. 
%Total revenue of firms are aggregated for each country and are shown in Fig.~\ref{fig:GDP_revenue}.
The aggregated revenue is compared with Gross Domestic Product (GDP) for each country as shown in Fig~\ref{fig:GDP_revenue} .  
This statistics provides an evidence for goodness of data coverage of our global supply chain data. The GDP data was collected from \url{https://data.worldbank.org/}, which is in public domain.

%\begin{table}[!h]
%\centering
%\caption{\bf Types of business relationship for non-US firms}
%\begin{tabular}{c|r|r}
%  \hline
%  Relationship & \#Links & Ratio (\%) \\
%  \hline 
%  Supplier & 587,369 & 64.1  \\
%  Creditor & 220,803 & 24.1  \\
%%  Transfer Agent & 2,615 & 0.3  \\
%%  Vendor & 7,710 & 0.8  \\
%%  Franchisor & 1,778 & 0.2  \\
%%  Investor Relations Firm & 84 & 0.0  \\
%  Landload & 50,414 & 5.5  \\
%%  Lessor	 & 23,868 & 2.6  \\
%  Licensor & 22,281 & 2.4  \\
%  \hline
%  Total links & 916,922 & 100 \\
%  \hline
%\end{tabular}
%\label{tab:Non-US}
%\end{table}

%\begin{table}[!h]
%\centering
%\caption{\bf Types of business relationship for US firms}
%\begin{tabular}{c|r|r}
%  \hline
%  Relationship & \#Links & Ratio (\%) \\
%  \hline 
%  Supplier & 261,854 & 45.8  \\
%  Creditor & 244,609 & 42.8  \\
%%  Transfer Agent & 491 & 0.1  \\
%%  Vendor & 1,892 & 0.3  \\
%%  Franchisor & 1,334 & 0.2  \\
%%  Investor Relations Firm & 43 & 0.0  \\
%  Landload & 26,494 & 4.6  \\
%%  Lessor	 & 9,663 & 1.7  \\
%  Licensor & 25,277 & 4.4  \\
%  \hline
%  Total links & 571,657 & 100 \\
%  \hline
%\end{tabular}
%\label{tab:US only}
%\end{table}

\begin{table}[!h]
\centering
\caption{\bf Types of business relationship for all firms}
\begin{tabular}{c|r|r}
  \hline
  Relationship & \#Links & Ratio (\%) \\
  \hline 
Supplier & 849,223 & 59.0  \\
Creditor & 465,412 & 32.3  \\
Landload & 76,908 & 5.3  \\
Licensor & 47,558 & 3.3  \\
\hline		
Total links & 1,439,101 & 100.0  \\
  \hline
\end{tabular}
\label{tab:Allfirms_link}
\end{table}

%\begin{table}[!h]
%\centering
%\caption{\bf Types of supplier for US firms}
%\begin{tabular}{c|r|r}
%  \hline
%  firm type & \#firms & Ratio (\%) \\
%  \hline 
%Private firm & 445,558 & 48.6 \\
%Private fund & 4,736 & 0.5 \\
%Private investment firm & 7,750 & 0.8 \\
%Public firm & 419,687 & 45.8 \\
%  \hline
%  Total firms & 916,922 & 100 \\
%  \hline
%\end{tabular}
%\label{tab:Non-US}
%\end{table}

%\begin{table}[!h]
%\centering
%\caption{\bf Types of supplier for US firms}
%\begin{tabular}{c|r|r}
%  \hline
%  firm type & \#firms & Ratio (\%) \\
%  \hline 
%Private firm & 385,357 & 67.4 \\
%Private fund & 27,444 & 4.8 \\
%Private investment firm & 12,414 & 2.2 \\
%Public firm & 114,223 & 20.0 \\
%  \hline
%  Total firms & 571,657 & 100 \\
%  \hline
%\end{tabular}
%\label{tab:US only}
%\end{table}

\begin{table}[!h]
\centering
\caption{\bf Types of supplier for all firms}
\begin{tabular}{c|r|r}
  \hline
  firm type & \#firms & Ratio (\%) \\
  \hline 
Private	firm & 830,915 & 58.6 \\
Private	fund & 32,180 & 2.3 \\
Private	investment firm & 20,164 & 1.4 \\
Public firm & 533,910 & 37.7 \\
\hline			
Total firms & 1,417,169 & 100.0 \\
\hline
\end{tabular}
\label{tab:Allfirms}
\end{table}

\section*{Methods}
\subsection*{Identification of bow-tie structure}

The bow-tie structure~\cite{broder2000graph} is uncovered from the GWCC based on the flow of goods and services (money flows in the opposite direction) along the directed links. The definitions of the different regions of the bow-tie structure are given as follows:
 \begin{itemize}
 \item The Giant strongly connected component (GSCC): The largest region where any two nodes are reachable through directed path. 
 \item IN components: The nodes from which GSCC is reachable through directed paths. 
 \item OUT components: The nodes that are reachable from the GSCC through directed paths.
 \item Tendrils (TE): The rest of the nodes in the GWCC. 
 \end{itemize}
We use breadth-first search algorithm to detect different components of bow-tie structure.

\subsection*{Community detection}

Empirical networks are generally non-homogeneous with a high local link density. 
Community detection captures highly connected groups of nodes as modules. It 
provides a coarse-grained description of  very large scale networks. 
Modularity maximization~\cite{newman2004fast} is one of the popular method to detect communities. 
In this method, one maximizes the modularity index. Modularity is defined as 
the fraction of intra-community links with a subtraction of the expected fraction given a random distribution. However, this method suffers from resolution limit problem~\cite{fortunato2007resolution} when applied to large networks. This indicate modularity optimization fails to detect well defined small communities in large scale networks. Moreover, this technique provides similar type of partition for both undirected and directed version of a network. It can not capture the dynamic behaviours of the network. 

The map equation method~\cite{rosvall2008maps} detects communities using the flow dynamics of the network. We use map equation method for our analysis as it is a directed network of suppliers and customers where link represents flow of goods.  This method is one of the best performing community detection techniques to detect communities in a network~\cite{lancichinetti2009community}. 
It minimizes per step average description length $L(C)$ of a random walker on the network as defined below
\begin{equation}
L(C)=q_\curvearrowright H(C)+\sum^m_{i=1}p^i_\circlearrowright H(\mathcal{P}^{i})\ .
%\label{mapeq}
\end{equation} 
$q_\curvearrowright$ and $H(C)$ are the probability and Shannon entropy for inter community movement of the random walker respectively.  
$p^i_\circlearrowright$ is the probability that the random walker leaves the node $i$, and  $H(\mathcal{P}^{i})$ is the entropy for intra community movement.

\subsection*{Over-expression of node attributes within communities}

Communities are ubiquitous in empirical networks. These communities are formed based on the similarities in some attributes of nodes. For examples, locations and sectors are key attributes for the formation of communities in Japanese supply chain network~\cite{chakraborty2018hierarchical, chakraborty2019characterization}, in protein-protein interaction networks, biological functions form the basis of community structure~\cite{chen2006detecting}.  

To measure the over-expression of attributes in a community we follow the method of
Tumminello {\em et. al.}~\cite{tumminello2011community}. In this method, the probability that $X$ randomly selected nodes in a community $C$ of size $N_C$
has the attribute $A$ is calculated by the following hyper-geometric distribution
$$H(X|N,N_C,N_A)=\frac{\binom{N_C}{X} \binom{N-N_C}{N_A-X}}{\binom{N}{N_A}},$$
where $N_A$ is the total number of nodes in the network with attribute $A$.
The $p$-value $p(N_{C,A})$ for the $N_{C,A}$ nodes with attribute $A$ in the community $C$
can be obtained from the following expression: 
$$p(N_{C,A})=1-\sum_{X=0}^{N_{C,A}-1}H(X|N,N_C,N_A).$$

The attribute $A$ is over-expressed when $p(N_{C,A})$ is lower than the some
threshold value $p_c$. As it is a multiple-hypothesis test, one has to choose
the $p_c$ appropriately to exclude false positive. We set $p_c=0.01/N_{A'}$ as
used in~\cite{tumminello2011community}, which takes care of the Bonferroni correction~\cite{miller1981normal}. Here,  $N_{A'}$ indicates total number of distinct attributes for all the nodes of the network.

\section*{Results}

\subsection*{Basic structural properties}
\begin{figure}[!h]
\centering
\includegraphics[width=\linewidth]{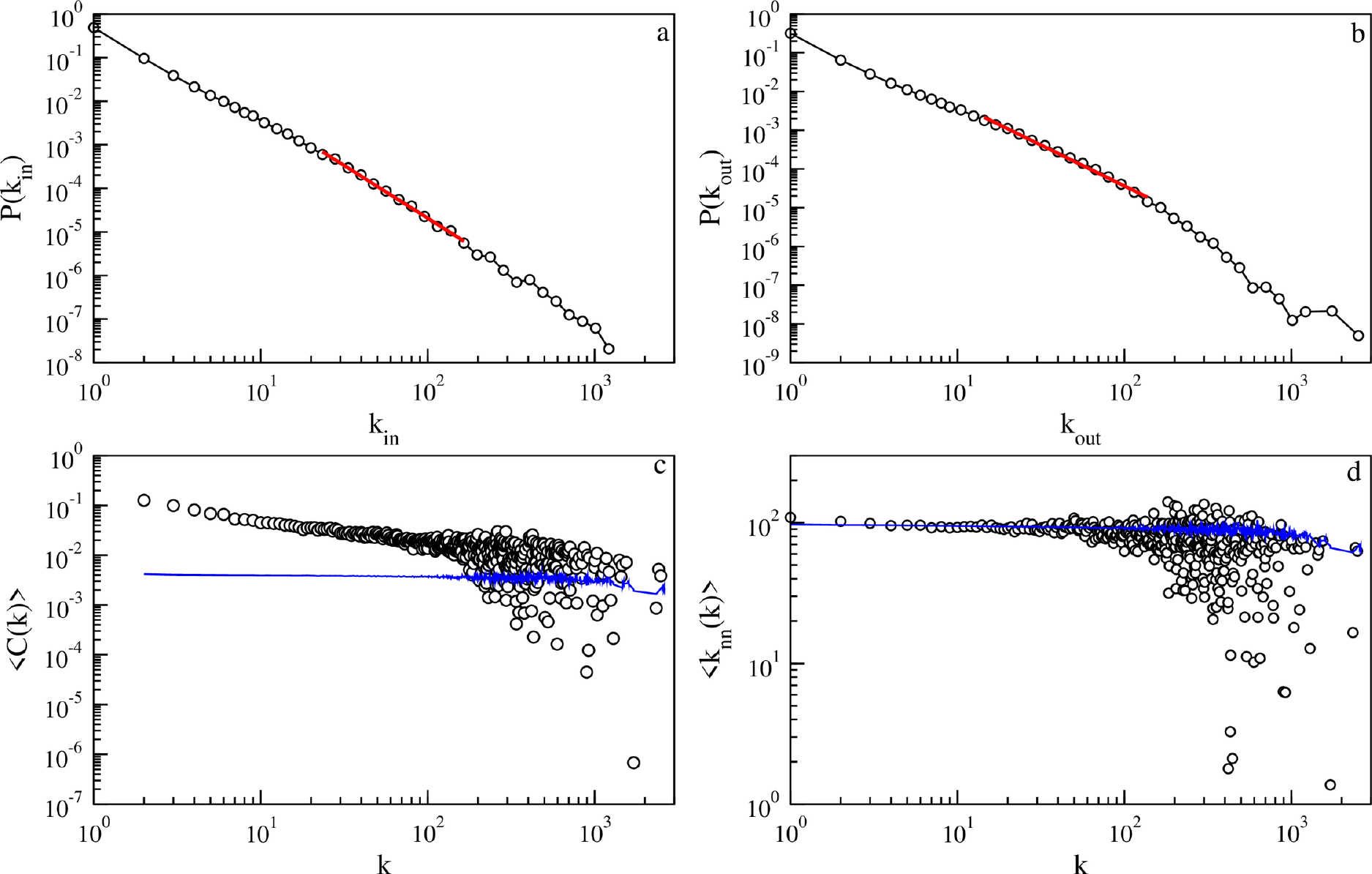}
\caption{{\bf Structural properties of the global supply chain network.
Probability density distributions $P$ of (a) the nodal in-degrees $k_{in}$ and (b) the nodal out-degrees $k_{out}$.
Variation of (c) the clustering coefficient $C(k) $ as a function of degree $k$
and (d) the average nearest neighbor degree $\langle k_{nn}(k) \rangle$ as a function of degree $k$.}
Logarithmic binning of the horizontal axis is used in (a) and (b). Red lines represent the best power-law fit to the data. Blue lines in (c) and (d) represent the results for degree preserved random network where average is taken over $100$ such uncorrelated networks.}
\label{fig:deg}
\end{figure}

As the supply chain network is directed in nature, one can define in and out degrees for the nodes. 
The nodal in-degree is defined as the number of incoming links to a node and out-degree is the total
number of outgoing links from that node.  We observe probability density distributions for both nodal in and out degree's have a heavy tail nature where 
the tail of the distributions is characterized by a power law of the form $P(k_{in/out}) \sim k^{-\gamma_{in/out}}$ with $\gamma_{in} =2.42$ and $\gamma_{out} = 2.11$ respectively as shown in  Fig.~\ref{fig:deg}~(a-b). The power law tail of the degree distribution is also observed in past investigations of empirical supply chain network data~\cite{sun2005scale, fujiwara2010large, mizuno2016structure, basole2016topological}. The degree distribution plays pivotal role in shock propagation among nodes. The high asymmetry in degree distribution can result in system wide aggregate fluctuation due to idiosyncratic shocks to large firms~\cite{acemoglu2012network}. It has been argued in the literature that such heavy tail distribution of nodal degrees arises due to rich-get-richer mechanism~\cite{barabasi1999emergence,chakraborty2010weighted}. Similar to the rich-get-richer principle, here large firms have more customers and suppliers than small firms. 

Clustering coefficient, a measure of three-point correlation, reflects cliquishness among the neighbours of a nodes. For most of the real world network, average clustering coefficient is a decaying function of degree having a form $\langle C(k) \rangle \sim k^{-\beta_k}$ with $\beta_k \leq 1.0$. We observe the clustering coefficient in the supply chain network decays with an exponent $\beta_k = 0.46$ as shown in Fig.~\ref{fig:deg}~(c) indicates the presence of a hierarchical structure. 

The average degree of the neighbors of a node $i$ which capture the nodal degree-degree correlation is defined as $k_{nn,i} = \sum_j k_j/k_i$ where the $j$ runs over all $k_i$ neighbours of $i$. For the nodes with degree $k$, $\langle k_{nn}(k) \rangle = \sum_{k_i=k} k_{nn,i}/N_k = \sum_{k_1}k_1 P(k_1|k)$ where $N_k$ is the number of nodes having degree $k$. The $\langle k_{nn}(k) \rangle$ increases with $k$ for a assoratative network and decreases for a disaasortative network. In the absence of nodal degree-degree correlation $\langle k_{nn}(k) \rangle$  remain constant. As can be seen from Fig. ~\ref{fig:deg}~(d),  $\langle k_{nn}(k) \rangle$ does not depend on $k$ and remain more or less in constant with $k$, indicating the absence of nodal degree-degree correlation. 
Further, the statistical significance of these results are tested by comparing it with 
results of the randomized degree preserving network~\cite{newman2010networks}.
The clustering coefficients of randomized network shows $C(k) \sim$ constant as expected. 
The variation of $\langle k_{nn}(k) \rangle$ with $k$ matches nicely with the case of degree preserving randomized network, which further supports the absence of nodal degree-degree correlation in the empirical network. 

\begin{figure}[!h]
\centering
\includegraphics[width=0.7\linewidth]{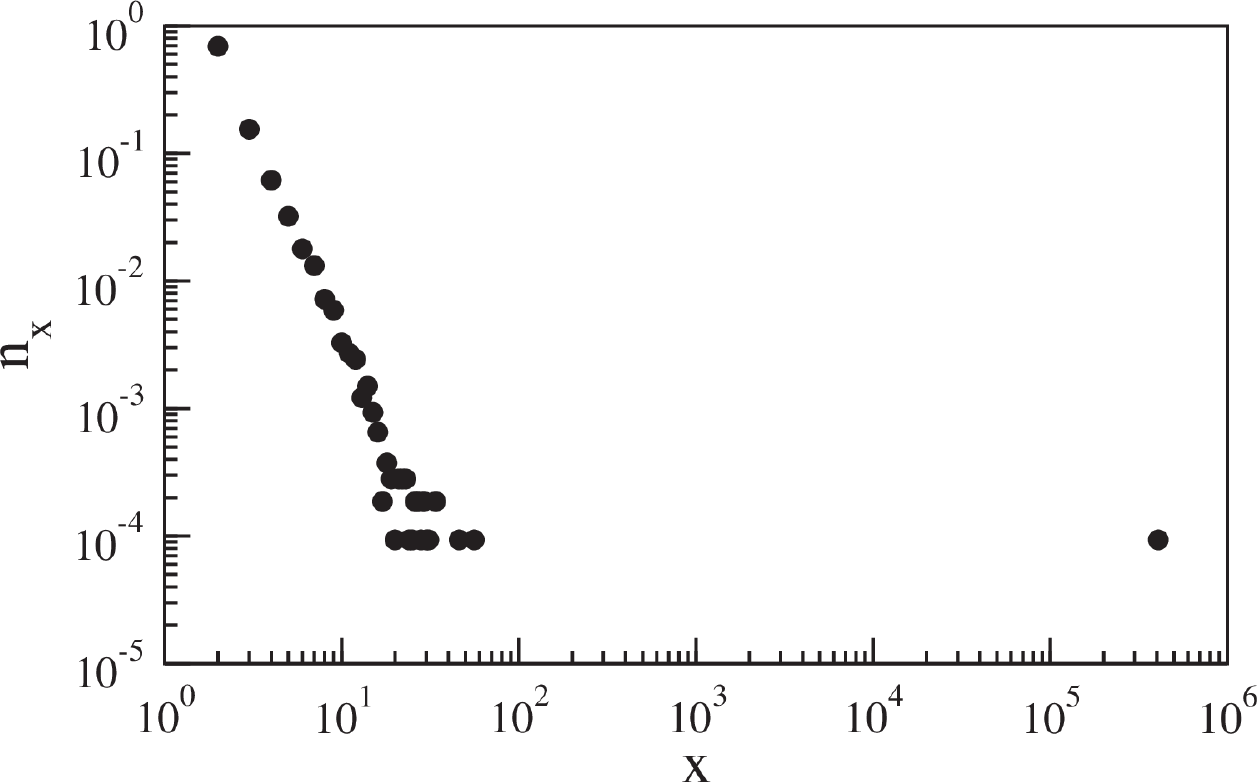}
\caption{{\bf Distribution $n_x$ of the component sizes $x$ in the network.
The largest weakly connected component contains $\sim 99\%$ of the nodes in the entire network.}
}
\label{fig:comp}
\end{figure}

We study the connected components when the network is viewed as an undirected network. The largest connected component of the network is known as the Giant weakly connected component(GWCC). As can be seen from Fig.~ \ref{fig:comp}, the network consists of a very large GWCC with $N = 407, 527$ nodes and $L = 927, 316$ links.  Using a breadth-first search, we calculate the average path length in the GWCC, by calculating the shortest paths between all pairs of nodes. The average path length is found to be $5.370$ reflecting the small world nature of the global supply chain network. While the GWCC contains $93.16\%$ of nodes of the network, rest of the components are very small.  In the subsequent sections, we investigate only the GWCC of the network. 

\subsection*{Bow-tie structure}
We detect the bow-tie components in GWCC of the global supply chain network. 
The number of firms in each component is shown in Table~\ref{tab:bowtie}. 
The OUT component, consists of nodes from which GSCC is reachable through directed paths toward downstream, is the largest and it consists $41.1\%$ of total firms.
GSCC (any two nodes are reachable through directed path), IN (nodes from which GSCC is reachable through directed paths toward upstream), and TE (The rest of the nodes in the GWCC) are approximately similar in size and comprise $16.4\%$, $22.3\%$, and $20.2\%$ of total firms respectively. 
For Japanese supply chain network, the fraction of each component of the OUT, GSCC, IN, and TE is $26.2\%$, $49.7\%$, $20.6\%$, and $3.5\%$, respectively~\cite{chakraborty2018hierarchical}.
The GSCC in the Japanese supply chain network occupies half of the system, meaning that most firms are interconnected by the small geodesic distances or the shortest-path lengths in the economy.
This shows a good contrast to the result of the global supply chain network observed in our study. 
However, by examining the shortest-path lengths from GSCC to IN and OUT as shown in
Table~\ref{tab:walnut_dist}, one can observe that the firms in the upstream or downstream sides are mostly located a few step away from the GSCC. This feature of the economic network is different from the bow-tie structure of many other complex networks~\cite{Broder2000}.

\begin{table}[!h]
\centering
\caption{\bf Bow-tie structure: Sizes of different components}
\begin{flushleft}
``Ratio" refers to the ratio of the number of firms to the total number of the firms in GWCC.
\end{flushleft}
\begin{tabular}{c|r|r}
  \hline
  Component & \#firms & Ratio (\%) \\
  \hline 
%  GWCC & 1,066,037 & 100\% \\   \hline
  GSCC & 66,798 & 16.4  \\
  IN & 90,992 & 22.3  \\
  OUT & 167,509 & 41.1  \\
  TE & 82,228 & 20.2  \\
  \hline
  Total & 407,527 & 100 \\
  \hline
\end{tabular}
\label{tab:bowtie}
\end{table}

\begin{table}[!h]
\centering
\caption{\bf Shortest distances from GSCC to IN/OUT}
\begin{tabular}[c]{crr|crr}
  \hline
  \multicolumn{3}{c|}{IN to GSCC} &
  \multicolumn{3}{c}{OUT to GSCC} \\ \hline
  Distance & \#firms&Ratio (\%)& Distance & \#firms &Ratio (\%)\\
  \hline
  1 & 82,761 &90.954&
  1 & 153,755 &91.789\\
  2 & 7,430 & 8.165&
  2 & 11,885 &7.095\\
  3 & 665 &0.731&
  3 & 1,582 &0.944\\
  4 & 104 &0.114&
  4 & 250 &0.149\\
  5 & 17 &0.019&
  5 & 26 &0.015\\
  6 & 10 &0.011&
  6 & 10 &0.006\\
  7 & 5 &0.005&
  7 & 1 &0.001\\
  \hline
  Total & 90, 992 &100&
  Total & 167, 509 &100\\
  \hline
\end{tabular}
\label{tab:walnut_dist}
\end{table}

\subsection*{Community structure}
\begin{figure}[!h]
\centering
\includegraphics[width=0.49\linewidth]{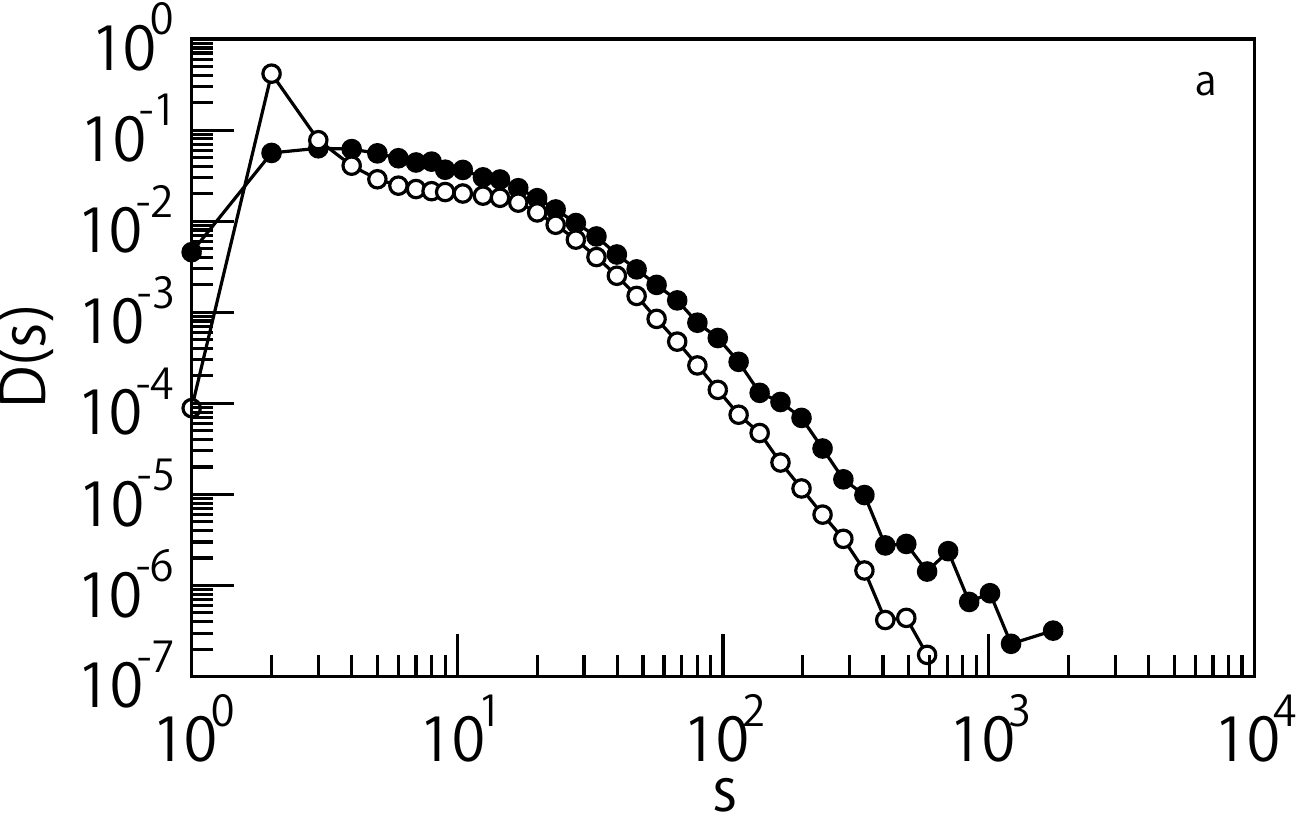}
\includegraphics[width=0.49\linewidth]{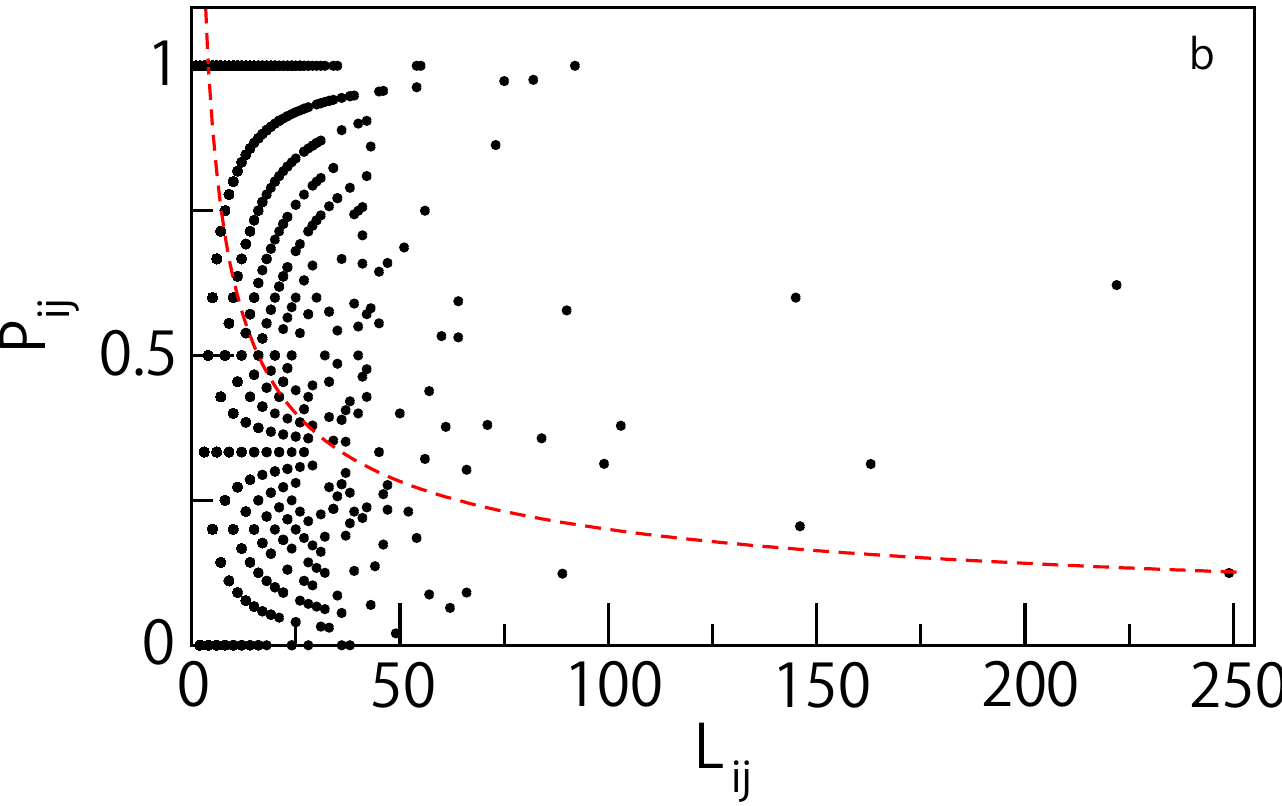}
\caption{{\bf Community sizes and polarizability} 
(a) Distributions $D(s)$ of community sizes $s$ for the actual network with directed links (filled circles) and for its randomized counterpart (open circles).
(b)Polarizability of the direction of the links interconnecting communities.
}
\label{fig:comsizepolarizability}
\end{figure}

Communities are detected in the largest weakly connected component of the network.
We employ the map equation method~\cite{rosvall2008maps} to uncover the communities in the GWCC of the global supply chain network.
The detected communities are found in various sizes. The probability density distributions $D(s)$ of community sizes $s$ for the empirical network and its degree preserving randomized network are shown in  Fig.~\ref{fig:comsizepolarizability} (a).
The distribution for the empirical network is more wider than it is for the randomized network. 

The biased in the direction of flow between a pair of communities is measured by the polarization ratio defined by $P_{ij}=|{w_{ij} - w_{ji}}|/(w_{ij} + w_{ji})$, where $w_{ij}$ is the total
number of links from $i$-th community to $j$-th community. $P_{ij} =1 $ if the flow is totally biased from one community to the other and $P_{ij} = 0$ if the flow is evenly balanced between 
the communities. The total flow between a pair of communities is $L_{ij} = (w_{ij} + w_{ji})$. If we assume that there is no bias in the flow direction between any pair of communities, according to a
null hypothesis, the values of $P_{ij}$  will fluctuates around $0$ with the standard deviation $\sigma=1/\sqrt{L_{ij}}$.  As can be seen from Fig.~\ref{fig:comsizepolarizability} (b), most of the values 
for the polarizability ratio $P_{ij}$ are significantly higher than the $2\sigma$ level which is indicated by the dashed curve. 

\subsection*{Overexpression within communities}
We study the significant overexpression of different attributes such as primary industry, sectors, firm's location, bow-tie components within the communities. 
We have shown the detail overexpression results within $10$ largest communities in Table~\ref{overexpression}. Various interesting features 
can be observed from the results of attribute overexpression. The largest community comprises of consumer discretionary sector based in the US. Further analysis shows these are private firms mainly from automotive retail, which
belong to the OUT component in the bow-tie structure of the global supply chain network. In the second largest community, we observe of consumer discretionary sector based in China, UK, France, 
Germany, Japan, Malaysia and New Zealand. These firms belong to the IN component of
bow-tie structure.  The firms of third largest communities are from consumer discretionary, industrials, and materials sectors which are mainly based in
Japan, China and Thailand. These firms are mostly belonged to TE component of bow-tie structure. 

\begin{table*}[!h] 
\begin {center}
 \scriptsize
\caption{Brief summary on the over-expression of sectors, countries and bow-tie components in the ten largest communities}
%\begin{tabular}{|l|l|p{4.0cm}|p{4.0cm}|p{2.5cm}|} \hline
\begin{tabular}{|l|l|p{3.5cm}|p{3.5cm}|p{2.5cm}|} \hline
  Rank   & Size       &  Sector  &  Country & Bow-tie components  \\ \hline
   1      &  1, 687    & Consumer Discretionary (26.7\%)  &  US (77.8\%) & OUT (99.7\%)         \\ 
  \hline
  2      &  1, 632 & Consumer Discretionary (40.4\%)  & China (9.1\%), UK (4.0\%), France (3.7\%), Germany (6.2\%), Japan (9.1\%), Malaysia (3.6\%), New Zealand (1.3\%) & IN (32.2\%)            \\
  \hline
  3      &  1, 179  & Consumer Discretionary (20.0\%), Industrials (23.1\%), Materials (15.8\%)&  
 China (12.4\%), Japan (53.1\%), Thailand (9.3\%) & TE (99.6\%)        \\
  \hline
  4      &  1, 027  & Communication Services (9.4\%)
  & Cambodia (0.6 \%), Indonesia (66.7\%), Singapore (3.9\%) & GSCC (20.3\%), IN (55.1\%)      \\ 
  \hline
  5      &   968  & Financials (62.1\%)     &  Bahrain(0.7\%), Bangladesh (1.8\%), Hong Kong (5.0\%), Hungary (1.0\%), Italy (4.0\%), 
Nepal (1.0\%), Pakistan (14.0\%), Singapore (3.8\%), UAE (2.4\%), Yemen (0.4\%) &   GSCC (22.7\%), IN (63.1\%)     \\                  
  \hline
  6      &  933  & Industrials (59.1\%) & 
US (93.3\%) &  IN (82.3\%)        \\                
  \hline
  7      &  867  & Information Technology (50.2\%)  & Korea South (3.1\%), Singapore (4.4\%), Taiwan (10.4\%)    &   GSCC (23.9\%), IN (29.6\%)        \\         
  \hline
  8      &  824  & Consumer Discretionary (20.0\%), Industrials (38.4\%) & US (52.4\%)    &   GSCC (29.5\%), IN (28.9\%)          \\ 
  \hline
  9      &  723  & Energy (12.3\%), Industrials (17.6\%), Materials (8.4\%), Utilities (7.5\%) & India(62.8\%)  &    GSCC (28.5\%)    \\
  \hline
  10     &  711  &  - &  Botswana (1\%), South Africa (69.5\%)      &  OUT (98.5\%)           \\ \hline
     \end{tabular}
\label{overexpression}
\end {center}
\end {table*}

\begin{figure}[!h]
\centering
\includegraphics[width=0.95\linewidth]{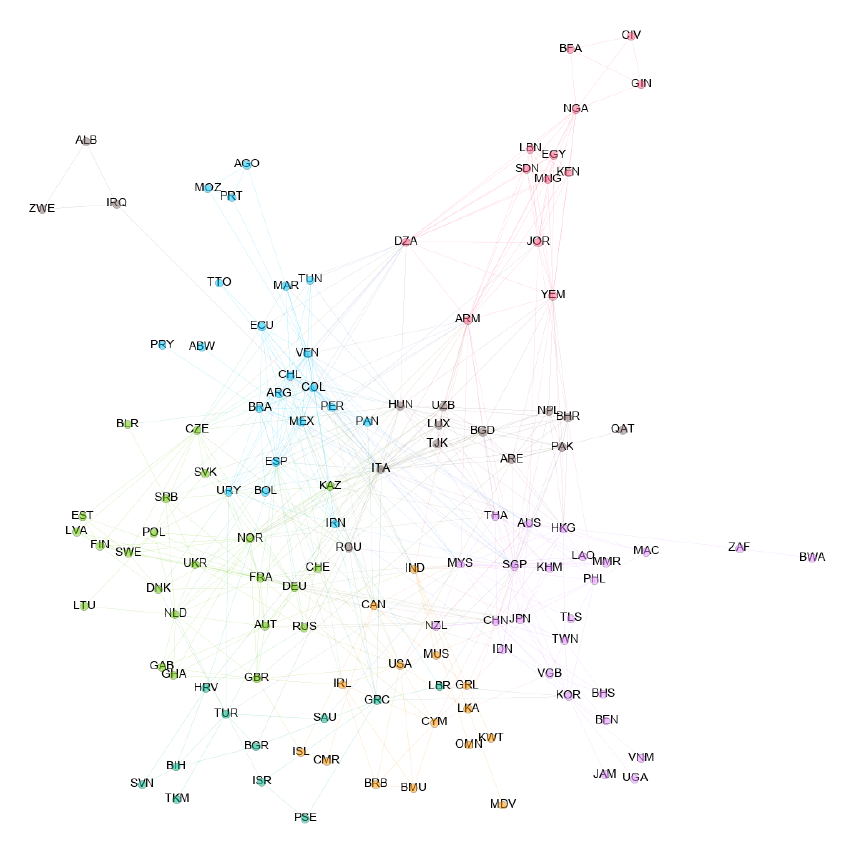}
\caption{{\bf  
Overexpression network of countries.} Different node color indicates different communities of the network.  Here, communities are detected using modularity maximization technique. 
}
\label{fig:country}
\end{figure}

We construct a weighted and undirected network of countries from their overexpression in communities with size larger than $100$ to show the inter-relation between countries. A link of weight $1$ is placed between two countries if they over-express simultaneously within a community.  Furthermore, we visualize community structure of this network as shown in Fig~\ref{fig:country}. It shows each community is formed by geographically closely located countries. 

Similarly, we also constructed a weighted undirected network of over-expressed primary industries, where a link of weight $w$ is present between two primary industries, if they are over-expressed simultaneously in $w$ communities. 
As can be seen from Fig~\ref{fig:sector} and Fig~S1 of Appendix S1, the clusters
among primary industries are formed based on their sector classification. 

\begin{figure}[!h]
\centering
\includegraphics[width=0.85\linewidth]{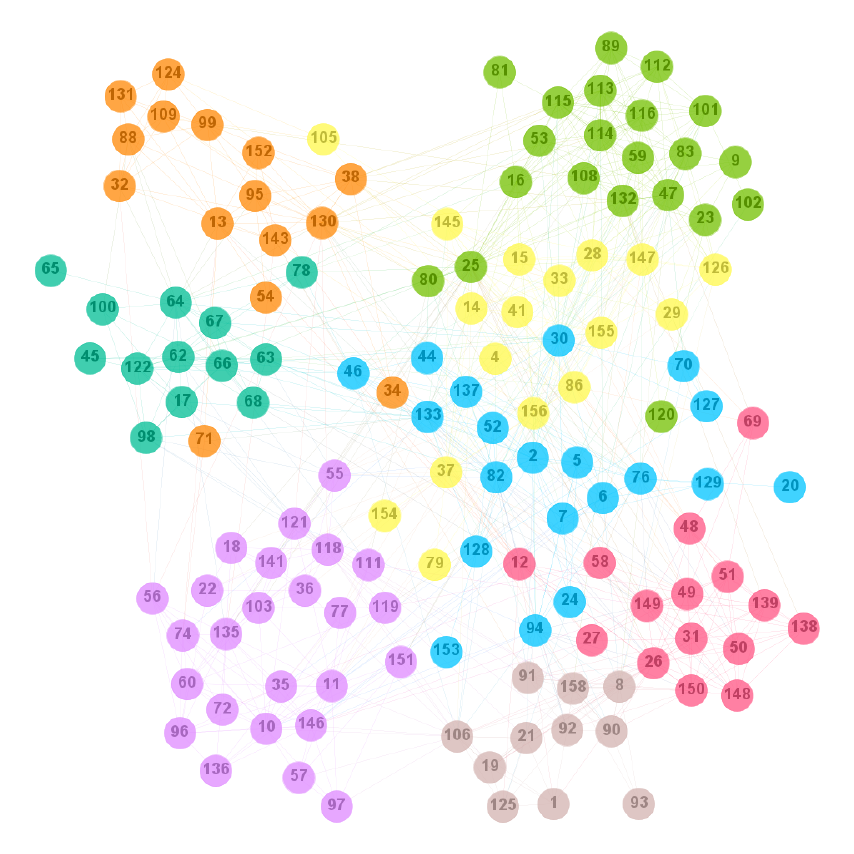}
\caption{{\bf  
Overexpression network of primary industries. Different node color indicates different communities of the network. Here, communities are detected using modularity maximization technique. IDs of the nodes are given in Table~S3 of Appendix S1.}
}
\label{fig:sector}
\end{figure}

\begin{figure}[!h]
\centering
\includegraphics[width=0.7\linewidth]{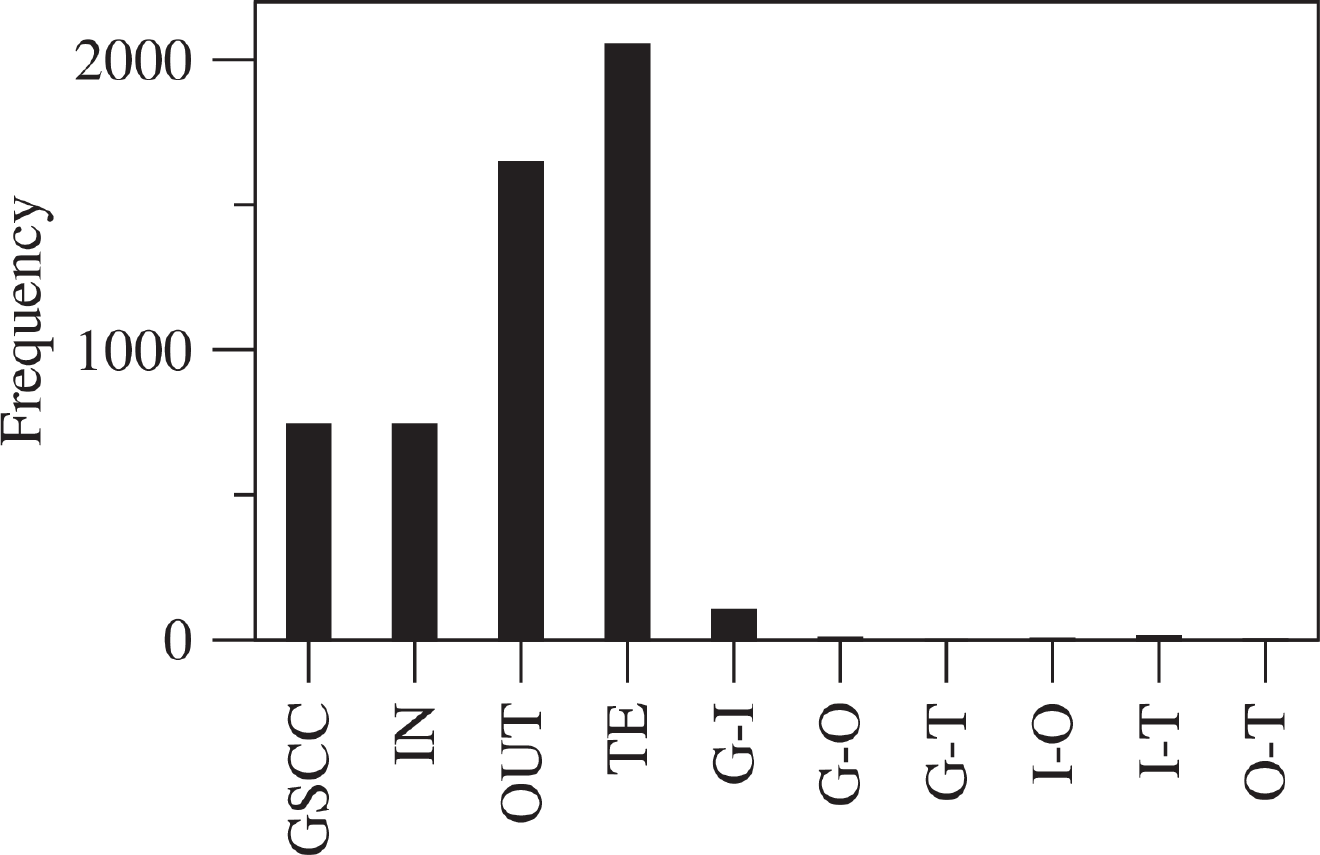}
\caption{{\bf Frequency of overexpression of bow-tie components within the communities with size at least $10$.} 
G-I indicates both GSCC and IN components are overexpressed in the communities. Similarly, G-O, G-T, I-O, I-T and O-T represent overexpression of GSCC-OUT, GSCC-TE, IN-OUT, IN-TE, and OUT-TE respectively. }
\label{fig:bowtie}
\end{figure}

We show the frequency of over-expression of the different components within the communities in bow-tie structure in Fig.~\ref{fig:bowtie}. 
Here, we selected communities which size of communities is at least $10$ firms. 
G-I indicates both GSCC and IN components are overexpressed in the communities. Similarly, G-O, G-T, I-O, I-T and O-T represent overexpression of GSCC-OUT, GSCC-TE, IN-OUT, IN-TE, and OUT-TE respectively. It reflects that most of the communities are solely composed of a particular component of the bow-tie structure. 
We also observe there are reasonable number of communities are composed by the combination
of GSCC and IN (G-I component), which is also observed in Japanese supply chain network~\cite{chakraborty2018hierarchical}. This indicate the flow of goods in the supply chain network is more often confined within the GSCC and IN component compared to any other combination of the components of bow-tie structure. 
Surprisingly, a large fraction of communities are located in TE component. The firms in the communities located in TE components not only supply but also procure any products and services from GSCC components.

\section*{Discussion on the nine propositions}
E.~J.S.~Hearnshaw~{\em et. al.}~\cite{Hearnshaw2013} have studied the supply chain network in terms of complex network approach and have proposed the nine propositions.
In this section, we investigate the validity of the nine propositions based on the obtained results on the topological properties of global supply chain network.

\subsection*{Proposition S1}
S1: \textit{Efficient supply chain systems demonstrate a short characteristic path length}

The average path length in the GWCC of the global supply chain was found to be $5.370$.
The average path length in the small world network $L_s$ is known to be similar to the average path length in the random graph $Ls \sim L_r$.
The average path length in the random graph $L_r$ is approximately calculated by $L_r = \log N/\log<k> = 6.77$. Here, the number of nodes in the GWCC is $N= 407527$, the average degree is $<k>= (<k_{in}> + <k_{out}>)/2 = 6.74$.
By assuming the degree distributions to be a power-law distributions in the entire range of the degree with $\gamma_{in} = 2.42$ and $\gamma_{out} = 2.11$.
The average in-degree and the average out-degree are calculated by 
$<k_{in}> = k_{in}^{min} (\gamma_{in}-1)/(\gamma_{in}-2) = 3.38$, and
$<k_{out}> = k_{out}^{min}(\gamma_{out}-1)/(\gamma_{out}-2) = 10.0$,
where $k_{in}^{min}=1$ and $k_{out}^{min}=1$.
The estimated value of $L_r= 6.77$ is close to the observed value $5.370$. 
This is reflecting the small world nature of the global supply chain network.  
Therefore the estimation of the average path length validate the proposition S1.

\subsection*{Propositions S2}
S2: \textit{The nodal degree distribution of efficient supply chain systems follows a power law as indicated by the presence of hub firms}

We observe probability density distributions for both nodal in and out degree's have a heavy tail nature where the tail of the distributions is characterized by a power law of the form $P(k_{in/out}) \sim k^{-\gamma_{in/out}}$ with $\gamma_{in} =2.42$ and $\gamma_{out} = 2.11$ respectively as shown in Fig.~\ref{fig:deg}~(a-b). 
The network whose degree distribution is characterized by a power law possess hub firms. The hub firms are known as channel leader firms which are said to control performance and provide system-wide coordination of the supply chain
\cite{Jarillo1988, Human2000}. The channel leader firms can exert their influence and provide opportunities and motivation for other firms to align themselves with their own specific objectives \cite{Cowan2007}. 
The power law distributions characterized with $\gamma_{in} = 2.42$ and $\gamma_{out} = 2.11$ validate the proposition S2.

\subsection*{Propositions S3}
S3: \textit{Efficient supply chain systems demonstrate a high clustering coefficient}

Clustering coefficient, a measure of three-point correlation, reflects cliquishness among the neighbours of a node. For most of the real world network, average clustering coefficient is a decaying function of degree having a form $\langle C(k) \rangle \sim k^{-\beta_k}$ with $\beta_k \leq 1.0$. We observe the clustering coefficient in the supply chain network decays with an exponent $\beta_k = 0.46$ as shown in Fig.~\ref{fig:deg}~(c) indicates the presence of a hierarchical structure. 
The observed moderate clustering coefficient indicates that the proposition S3: \textit{It has a high clustering coefficient} is weakly valid.

\subsection*{Propositions S4}
S4: \textit{The growth of efficient supply chain systems follows “fit-gets-richer" mechanism}

It has been argued in the literature that such heavy tail distribution of nodal degrees arises due to rich-get-richer mechanism\cite{barabasi1999emergence,chakraborty2010weighted}. Similar to the rich-get-richer principle, here large firms have more customers and suppliers than small firms. 
Preferential attachment in rich-get-richer mechanism assumes that the acquisition of new links by a firm is determined solely by the number of its existing links. This assumption leads to the number of links being proportional to their duration in the supply chain. However, one can often observe that older firms have been outstripped by new entrant firms.
There is a need therefore, to include the “fitness” of the firms to account for new entrants that can quickly dominate supply chains.
By introducing “fit-gets-richer” mechanism \cite{Bianconi2001}, 
the fitter nodes have a greater acquisition rate for links and therefore, resulting network possess a scale-free property.
The heavy tail distribution of nodal degrees and overtaking of older firms by new entrant firms validate the proposition S4.
 
\subsection*{Propositions S5}
S5: \textit{The power law degree distribution of efficient supply chain system is truncated}

The power law distributions $P(k_{in/out}) \sim k^{-\gamma_{in/out}}$ with $\gamma_{in} =2.42$ and $\gamma_{out} = 2.11$ respectively are observed in the middle region of the distributions as shown in Fig.~\ref{fig:deg}~(a-b).
The tail region of both distributions seem like truncated or exponentially cut-off.   
Especially this tendency is evident for $P(k_{out})$.
This phenomenon is said to be caused by four reasons \cite{Hearnshaw2013}.
First, the finite size of marketplaces generates a truncated power law degree distribution. 
Second, there are practical reasons in the operation of firms that limit the ability of firms to indefinitely form and maintain exchange relationships.
Third, when new links are to be formed with a hub firms, incomplete information generates uncertainty which might costs higher than transaction costs. If these costs are unacceptable, the firms will scrap the deal with the hub firms. 
Finally, the aging and depreciation of firms limits their growth. 
The observed truncation or cut-off in the tail region of the degree distribution validates the proposition S5.

\subsection*{Propositions S7}
S7: \textit{Efficient supply chain systems demonstrate a pronounced community structure with overlapping boundaries}

We employ the map equation method~\cite{rosvall2008maps} to uncover the communities in the GWCC of the global supply chain network.
The detected communities are found in various sizes. The probability density distributions $D(s)$ of community sizes $s$ for the empirical network and its degree preserving randomized network are shown in Fig~\ref{fig:comsizepolarizability} (a).
The distribution for the empirical network is more wider than it is for the randomized network. 
In Table~\ref{overexpression}, the over-expression of sectors and countries in the ten largest communities is shown.
Communities in a supply chain are bound together in clusters predominantly connected by horizontal relationships among firms with similar interests and functions. However, we empirically observed that all firms within
a community are not entirely cooperative as shown in Table~\ref{overexpression}. Therefore, community formation in supply chain possess overlapping boundaries.
These results validates the proposition S7.

\subsection*{Remaining Propositions}
The supply chain data has no weight on links. Therefore the following two hypotheses:
S6: \textit{The link weight distribution of efficient supply chain systems follows a power law}, 
S9: \textit{Resilient supply chain systems demonstrate a power law distribution for link-weights}
are not applicable in the analyses of this paper.
In addition, we concentrated on the topological properties of the supply chain network and therefore, the resilience of the system:
S8: \textit{The fitness of hub firms determines the resilience of supply chain systems against both random disturbances and targeted attacks} 
is out of scope of our current study.

\section*{Conclusions}
We studied on topological properties of global supply chain network in terms of degree distribution, hierarchical structure, and degree-degree correlation in the global supply chain network. 
The global supply chain data was constructed by collecting various company data from the website of Standard \& Poor's Capital IQ platform in 2018.
The total number of firms and directed links in our data were $437,453$ and $948,247$, respectively.

The degree distributions is characterized by a power law of the form with $\gamma_{in} = 2.42$ and $\gamma_{out} = 2.11$. The clustering coefficient decays $\langle C(k) \rangle \sim k^{-\beta_k}$ with an exponent $\beta_k = 0.46$. This indicates the presence of a hierarchical structure of the supply chain network. We observed that $\langle k_{nn}(k) \rangle$ does not depend on $k$ and remain more or less in constant with $k$, indicating the absence of nodal degree-degree correlation. The Bow-tie structure of GWCC revealed that the OUT component was the largest and it consists $41.1\%$ of total firms. The GSCC component comprised $16.4\%$ of total firms. We observed that the firms in the upstream or downstream sides were mostly located a few step away from the GSCC. 

Furthermore, we uncovered the community structure of the network using map equation method and characterized them according to their location and industry classification.  
We observed that the largest community comprises of private firms mainly from automotive retail based in the US. These firms are belong to the OUT component in the bow-tie structure of the global supply chain network. It indicates the retail firms are generally belong to the OUT component of bow-ties structure.

Finally, we investigated the validity of the nine propositions on the supply chain network based on the obtained results on the topological properties.
We confirmed the validity of propositions S1 (short path length), S2 (power-law degree distribution), S3 (high clustering coefficient), S4 (“fit-gets-richer” growth mechanism), S5 (truncation of power-law degree distribution), and S7 (community structure with overlapping boundaries) in the global supply chain network. However, the propositions related to link weight and resilient nature of the network were not confirmed due to the limitation of our data and the scope of our current study. This will be left for future study.

Our study provides a detailed topological characterization of the global supply chain network. These topological properties are utmost important to understand the international trade dynamics. 
It is well-known that community structure plays an important role in spreading phenomena. Our characterization of community structure will be helpful to understand the wide-spread economic crisis. 
The study further shows the inter-relationships among the countries and among the industrial sectors.

\section*{Acknowledgments}
We are grateful to Y. Fujiwara, H. Aoyama, H. Iyetomi, W. Souma, and H. Yoshikawa for their insightful comments and encouragement.
The present study was supported by the Ministry of Education, Science, Sports, and Culture,
Grants-in-Aid for Scientific Research (B), Grant no. 17KT0034 (2017-2019) and Exploratory Challenges
on Post-K computer (Studies of Multi-level Spatiotemporal Simulation of Socioeconomic Phenomena).

%\nolinenumbers

% Either type in your references using
% \begin{thebibliography}{}
% Text
% \end{thebibliography}
%
% or
%
% Compile your BiBTeX database using our plos2015.bst
% style file and paste the contents of your .bbl file
% here. See http://journals.plos.org/plosone/s/latex for 
%\bibliography{sample}

\begin{thebibliography}{10}

\bibitem{Bellamy2012}
Bellamy MA, Basole RC.
\newblock Network analysis of supply chain systems: A systematic review and
  future research.
\newblock Systems Engineering. 2012;16(2):235--249.
\newblock doi:{10.1002/sys.21238}.

\bibitem{Perera2017}
Perera S, Bell MGH, Bliemer MCJ.
\newblock Network science approach to modelling the topology and robustness of
  supply chain networks: a review and perspective.
\newblock Applied Network Science. 2017;2(1).
\newblock doi:{10.1007/s41109-017-0053-0}.

\bibitem{Bell2017}
Bell M, Perera S, Piraveenan M, Bliemer M, Latty T, Reid C.
\newblock Network growth models: A behavioural basis for attachment
  proportional to fitness.
\newblock Scientific Reports. 2017;7(1).
\newblock doi:{10.1038/srep42431}.

\bibitem{Econophysics2010}
Fujiwara Y, Ikeda Y.
\newblock 6.
\newblock In: Econophysics and Companies -Statistical Life and Death in Complex
  Business Networks-. Cambridge CB2 8RU, UK: Cambridge University Press; 2010.
  p. 184--220.

\bibitem{Mizgier2012}
Mizgier KJ, Wagner SM, Holyst JA.
\newblock Modeling defaults of companies in multi-stage supply chain networks.
\newblock International Journal of Production Economics. 2012;135(1):14--23.
\newblock doi:{10.1016/j.ijpe.2010.09.022}.

\bibitem{Tang2016}
Tang L, Jing K, He J, Stanley HE.
\newblock Complex interdependent supply chain networks: Cascading failure and
  robustness.
\newblock Physica A: Statistical Mechanics and its Applications.
  2016;443:58--69.
\newblock doi:{10.1016/j.physa.2015.09.082}.

\bibitem{mizuno2016structure}
Mizuno T, Ohnishi T, Watanabe T.
\newblock Structure of global buyer-supplier networks and its implications for
  conflict minerals regulations.
\newblock EPJ Data Science. 2016;5(1):2.

\bibitem{clauset2004finding}
Clauset A, Newman ME, Moore C.
\newblock Finding community structure in very large networks.
\newblock Physical review E. 2004;70(6):066111.

\bibitem{Hearnshaw2013}
Hearnshaw EJS, Wilson MMJ.
\newblock A complex network approach to supply chain network theory.
\newblock International Journal of Operations {\&} Production Management.
  2013;33(4):442--469.
\newblock doi:{10.1108/01443571311307343}.

\bibitem{broder2000graph}
Broder A, Kumar R, Maghoul F, Raghavan P, Rajagopalan S, Stata R, et~al.
\newblock Graph structure in the web.
\newblock Computer networks. 2000;33(1-6):309--320.

\bibitem{newman2004fast}
Newman ME.
\newblock Fast algorithm for detecting community structure in networks.
\newblock Physical review E. 2004;69(6):066133.

\bibitem{fortunato2007resolution}
Fortunato S, Barthelemy M.
\newblock Resolution limit in community detection.
\newblock Proceedings of the national academy of sciences. 2007;104(1):36--41.

\bibitem{rosvall2008maps}
Rosvall M, Bergstrom CT.
\newblock Maps of random walks on complex networks reveal community structure.
\newblock Proceedings of the National Academy of Sciences.
  2008;105(4):1118--1123.

\bibitem{lancichinetti2009community}
Lancichinetti A, Fortunato S.
\newblock Community detection algorithms: a comparative analysis.
\newblock Physical review E. 2009;80(5):056117.

\bibitem{chakraborty2018hierarchical}
Chakraborty A, Kichikawa Y, Iino T, Iyetomi H, Inoue H, Fujiwara Y, et~al.
\newblock Hierarchical communities in the walnut structure of the Japanese
  production network.
\newblock PloS one. 2018;13(8).

\bibitem{chakraborty2019characterization}
Chakraborty A, Krichene H, Inoue H, Fujiwara Y.
\newblock Characterization of the community structure in a large-scale
  production network in Japan.
\newblock Physica A: Statistical Mechanics and its Applications.
  2019;513:210--221.

\bibitem{chen2006detecting}
Chen J, Yuan B.
\newblock Detecting functional modules in the yeast protein--protein
  interaction network.
\newblock Bioinformatics. 2006;22(18):2283--2290.

\bibitem{tumminello2011community}
Tumminello M, Micciche S, Lillo F, Varho J, Piilo J, Mantegna RN.
\newblock Community characterization of heterogeneous complex systems.
\newblock Journal of Statistical Mechanics: Theory and Experiment.
  2011;2011(01):P01019.

\bibitem{miller1981normal}
Miller RG.
\newblock Normal univariate techniques.
\newblock In: Simultaneous statistical inference. Springer; 1981. p. 37--108.

\bibitem{sun2005scale}
Sun H, Wu J.
\newblock Scale-free characteristics of supply chain distribution networks.
\newblock Modern physics letters B. 2005;19(17):841--848.

\bibitem{fujiwara2010large}
Fujiwara Y, Aoyama H.
\newblock Large-scale structure of a nation-wide production network.
\newblock The European Physical Journal B. 2010;77(4):565--580.

\bibitem{basole2016topological}
Basole RC.
\newblock Topological analysis and visualization of interfirm collaboration
  networks in the electronics industry.
\newblock Decision Support Systems. 2016;83:22--31.

\bibitem{acemoglu2012network}
Acemoglu D, Carvalho VM, Ozdaglar A, Tahbaz-Salehi A.
\newblock The network origins of aggregate fluctuations.
\newblock Econometrica. 2012;80(5):1977--2016.

\bibitem{barabasi1999emergence}
Barab{\'a}si AL, Albert R.
\newblock Emergence of scaling in random networks.
\newblock science. 1999;286(5439):509--512.

\bibitem{chakraborty2010weighted}
Chakraborty A, Manna S.
\newblock Weighted trade network in a model of preferential bipartite
  transactions.
\newblock Physical Review E. 2010;81(1):016111.

\bibitem{newman2010networks}
Newman M.
\newblock Networks: an introduction.
\newblock Oxford university press; 2010.

\bibitem{Broder2000}
Broder A, Kumar R, Maghoul F, Raghavan P, Rajagopalan S, Stata R, et~al.
\newblock Graph structure in the web.
\newblock Computer networks. 2000;33(1-6):309--320.

\bibitem{Jarillo1988}
Jarillo C.
\newblock On strategic networks.
\newblock Strategic Management Journal. 1988;9(1):31--41.

\bibitem{Human2000}
Human SE, Provan KG.
\newblock Legitimacy building in the evolution of small firm multilateral
  networks: a comparative study of success and demise.
\newblock Administrative Science Quarterly. 2000;45(2):327--65.

\bibitem{Cowan2007}
Cowan R, Jonard N.
\newblock Structural holes, innovation and the distribution of ideas.
\newblock Journal of Economic Interaction and Coordination. 2007;2(2):93--110.

\bibitem{Bianconi2001}
Bianconi G, Barab{\'a}si AL.
\newblock Competition and multiscaling in evolving networks.
\newblock EPL (Europhysics Letters). 2001;54(4):436.

\end{thebibliography}
%\begin{thebibliography}{10}

\section*{Supporting information}

\paragraph*{S1 Appendix.}
\label{S1_Appendix}
{\bf Appendix to the manuscript.} 

\newcommand{\beginsupplement}{%
        \setcounter{table}{0}
        \renewcommand{\thetable}{\Alph{table}}
        \setcounter{figure}{0}
        \renewcommand{\thefigure}{\Alph{figure}}
     }
\newcommand{\eq}[1]{Eq.~(\ref{#1})}
\newcommand{\fig}[1]{Fig.~\ref{#1}}

\newcommand{\dts}{\partial_t}
\newcommand{\dxs}{\partial_x}
\newcommand{\dvs}{\partial_v}

%==================================================================================================%
\beginsupplement     
\title{Appendix S1: Bow-tie structure and community identification of global supply chain network}

%\author{Shakti N. Menon,~Trilochan Bagarti and Abhijit Chakraborty}
%\affiliation{
%The Institute of Mathematical Sciences, CIT Campus, Taramani, Chennai 600113, India
%}

\maketitle

\begin{itemize}
\item Table~\ref{tab:country} represents the code of countries, firm distribution in different countries, total revenue of the firms in each country and countries' GDP (current US\$) for the year $2017$.
\item Table~\ref{tab:sectors} represents the industrial sectors and firm distribution. 
\item Table~\ref{tab:primaryindustry} shows the primary industry and sector classification. 

\end{itemize}

%-------------------------------------------
{\tiny
\begin{longtable}{|r|l|c|r|r|r|}
\hline
Sl. no.&Country name& Code& \#firms&Total revenue&GDP (current US\$) 2017  \\
\hline
1 & Afghanistan &AFG & 34 & 365.6 & 20191764940.1602 \\
2&Albania&ALB&64&1038.8&13025062195.7906 \\
3&Algeria&DZA&135&16686.7&167555280113.181 \\
4&Andorra&AND&8&464.53&3013387423.93509 \\
5&Angola&AGO&132&27427.51&122123822333.591 \\
6&Anguilla&AIA&7&0.279& - \\
7&Antigua and Barbuda&ATG&23&-&1510084750.74074 \\
8&Argentina&ARG&1199&178835.56&642695864756.35 \\
9&Armenia&ARM&56&1241.83&11527458565.7334 \\
10&Aruba&ABW&19&238.7&2700558659.21788 \\
11&Australia&AUS&11955&3894482.47199999&1330803227996.08 \\
12&Austria&AUT&1500&495115.007&416835975862.194 \\
13&Azerbaijan&AZE&83&32534.556&40865558912.3867 \\
14&Bahamas, The&BHS&131&1036.314&12162100000 \\
15&Bahrain&BHR&249&19700.806&35432686170.2128 \\
16&Bangladesh&BGD&1514&29319.103&249723862487.361 \\
17&Barbados&BRB&104&3406.7&4673500000 \\
18&Belarus&BLR&128&8661.46&54726595249.1849 \\
19&Belgium&BEL&2296&503814.929&494901708704.269 \\
20&Benin&BEN&14&58.4&9246696923.66155 \\
21&Bermuda&BMU&677&215629.304&-\\
22&Bhutan&BTN&25&376.8&2528007911.35353 \\
23&Bolivia&BOL&108&11963.02&37508642257.5977 \\
24&Bonaire&BES&1&-&- \\
25&Bosnia and Herzegovina&BIH&112&3998.21&18080118128.3854 \\
26&Botswana&BWA&195&14360.19&17406565823.2986 \\
27&Brazil&BRA&4059&1341430.151&2053594973992.61 \\
28&British Virgin Islands&VGB&571&2646.326&-\\
29&Brunei&BRN&48&336&12128088999.9276 \\
30&Bulgaria&BGR&578&28528.031&58220973782.7715 \\
31&Burkina Faso&BFA&19&-&12322864244.9183 \\
32&Burundi&BDI&8&137.14&3172416146.3921 \\
33&Cote d'Ivoire&CIV&43&5911.3&38053610009.4172 \\
34&Cambodia&KHM&159&2191.8&22177200511.5811 \\
35&Cameroon&CMR&51&19648.33&34922782310.6416 \\
36&Canada&CAN&15016&2789244.548&1646867220617.47 \\
37&Cape Verde&CPV&9&173.5&1771235958.20125 \\
38&Cayman Islands&CYM&482&17415.518&3570575151.26723 \\
39&Central African Republic&CAF&4&-&2167501639.76783 \\
40&Chad&TCD&10&-&9975692095.40855 \\
41&Channel Islands&CHI&345&27779.03&-\\
42&Chile&CHL&1794&269727.3&277746457909.868 \\
43&China&CHN&27633&11960538.449&12143491448186.1 \\
44&Colombia&COL&1026&195727.402&311789874617.096 \\
45&Comoros&COM&3&-&1068124329.86257 \\
46&Congo (Brazzaville)&COG&49&1207.5&8701334800.21976 \\
47&Congo Democratic Republic of&COD&46&1412.7&38019265625.856 \\
48&Cook Islands&COK&5&-&- \\
49&Croatia&HRV&553&34162.332&55201417479.3925 \\
50&Cuba&CUB&17&-&96851000000 \\
51&Cyprus&CYP&538&58964.634&22141864998.8731 \\
52&Czech Republic&CZE&1372&145468.25&215913545038.43 \\
53&Denmark&DNK&1926&448918.056&329865537183.47 \\
54&Djibouti&DJI&3&154.9&1844674434.50371 \\
55&Dominica&DMA&11&-&496726248.518519 \\
56&Dominican Republic&DOM&77&3072.6&75931656814.657 \\
57&East Timor&TLS&8&-&2487269437.36822 \\
58&Ecuador&ECU&172&7090.82&104295862000 \\
59&Egypt&EGY&603&31545.301&235369129337.711 \\
60&Eritrea&ERI&4&-&-\\
61&Estonia&EST&370&13558.41&26611651598.9453 \\
62&Ethiopia&ETH&56&3458.8&81716326730.819 \\
63&Falkland Islands&FLK&4&-&- \\
64&Fiji&FJI&53&1195.54&5270335184.70768 \\
65&Finland&FIN&1920&340265.239&252301837573.029 \\
66&France&FRA&8165&3448556.89&2586285406561.51 \\
67&French Guiana&GUF&5&25.8&- \\
68&French Polynesia&PYF&7&69.7&-\\
69&Gabon&GAB&31&5759.1&14892609693.1667 \\
70&Gambia The&GMB&8&-&1489464787.85603 \\
71&Georgia&GEO&74&578.9&15081330942.4188 \\
72&Germany&DEU&11574&4555411.97100001&3693204332229.78 \\
73&Ghana&GHA&188&5918.08&58996776244.424 \\
74&Gibraltar&GIB&76&1307.06&-\\
75&Greece&GRC&1985&113381.706&203085551429.132 \\
76&Greenland&GRL&10&648.08&-\\\
77&Grenada&GRD&5&70.4&1126882296.2963 \\
78&Guadeloupe&GLP&5&-&- \\
79&Guinea&GIN&28&-&9915311049.15133 \\
80&Guinea-Bissau&GNB&4&3080.4&1346841897.00437 \\
81&Guyana&GUY&18&23&3555205811.13802 \\
82&Haiti&HTI&11&49.3&8408252995.16143 \\
83&Hong Kong&HKG&6725&1977552.828&341648103474.824 \\
84&Hungary&HUN&764&90165.09&139761138102.757 \\
85&Iceland&ISL&163&11710.79&24489493459.0074 \\
86&India&IND&18502&1675896.42300001&2652551202555.27 \\
87&Indonesia&IDN&10403&402879.001&1015423455783.28 \\
88&Iran&IRN&286&40698.59&454012768723.589 \\
89&Iraq&IRQ&93&1633.6&193158783783.784 \\
90&Ireland&IRL&1755&614668.552999999&331430014003.047 \\
91&Israel&ISR&2328&280344.674&353268411918.577 \\
92&Italy&ITA&6998&1371163.738&1946570340341.13 \\
93&Jamaica&JAM&191&7091.264&14781107821.7513 \\
94&Japan&JPN&10775&8675056.03899996&4859950558538.97 \\
95&Jordan&JOR&394&32611.639&40765867418.8999 \\
96&Kazakhstan&KAZ&426&35509.18&162886867831.694 \\
97&Kenya&KEN&348&21157.894&78757391333.0088 \\
98&Kiribati&KIR&1&-&185572501.532802 \\
99&Korea North&PRK&5&-&-\\\
100&Korea South&KOR&4997&2626491.358&1530750923148.7 \\
101&Kosovo&XKX&18&304.5&7227764976.79274 \\
102&Kuwait&KWT&351&43292.112&119551599076.822 \\
103&Kyrgyzstan&KGZ&35&452.2&7702934800.12836 \\
104&Laos&LAO&50&848.4&16853087485.4118 \\
105&Latvia&LVA&312&10913.567&30463302413.7289 \\
106&Lebanon&LBN&166&6849.5&53393799668.325 \\
107&Lesotho&LSO&13&4389.7&2578265355.71255 \\
108&Liberia&LBR&112&-&3285455000 \\
109&Libya&LBY&74&212.6&38115981878.5647 \\
110&Liechtenstein&LIE&43&12177.55&-\\\
111&Lithuania&LTU&524&25803.34&47544459558.9514 \\
112&Luxembourg&LUX&653&206746.656&62316359824.1281 \\
113&Macau&MAC&195&31353.683&50559431846.4989 \\
114&Macedonia&MKD&66&1299.02&11279509013.9119 \\
115&Madagascar&MDG&39&34.5&11465850504.0067 \\
116&Malawi&MWI&53&5275.6&6303292264.18905 \\
117&Malaysia&MYS&7936&408028.655000001&314707268049.991 \\
118&Maldives&MDV&43&356.45&4865546025.86599 \\
119&Mali&MLI&28&65.1&15339614406.6617 \\
120&Malta&MLT&271&10381.83&12748803180.3035 \\
121&Martinique&MTQ&5&-&- \\
122&Mauritania&MRT&17&727.3&4975432190.51025 \\
123&Mauritius&MUS&369&11615.152&13259351418.4459 \\
124&Mexico&MEX&2776&835615.566000002&1158071006809.62 \\
125&Moldova&MDA&72&1166.43&9669759987.02633 \\
126&Monaco&MCO&82&3054&6400946585.53076 \\
127&Mongolia&MNG&118&-418.7&11433635875.9316 \\
128&Montenegro&MNE&44&983.9&4844592066.71174 \\
129&Morocco&MAR&231&52080.09&109708728848.535 \\
130&Mozambique&MOZ&88&3827.3&12651912500.4128 \\
131&Myanmar&MMR&116&669.2&66719084835.9898 \\
132&Namibia&NAM&164&13079.7&13566192142.5835 \\
133&Nauru&NRU&2&-&113880715.219337 \\
134&Nepal&NPL&65&1420.2&24880266905.4961 \\
135&Netherlands&NLD&4149&1755592.252&830572618849.829 \\
136&Netherlands Antilles&ANT&58&8965.1&- \\
137&New Caledonia&NCL&15&13144.4&-\\\
138&New Zealand&NZL&2155&396634.359999999&202590814084.991 \\
139&Niger&NER&14&-&8119710126.32655 \\
140&Nigeria&NGA&1294&33653.433&375745486520.656 \\
141&Norway&NOR&3034&509958.53&399488897844.046 \\
142&Oman&OMN&492&27061.932&70783875162.5488 \\
143&Pakistan&PAK&1299&77117.213&304951818494.066 \\
144&Palestinian Authority&PSE&35&1998.84&14498100000 \\
145&Panama&PAN&150&15671.8&62283800000 \\
146&Papua New Guinea&PNG&78&21203.77&22277692408.8879 \\
147&Paraguay&PRY&67&1827.3&39008900331.6733 \\
148&Peru&PER&813&105205.9&210702303186.432 \\
149&Philippines&PHL&3255&242776.568&313619747740.186 \\
150&Poland&POL&5339&490258.466&526371021088.561 \\
151&Portugal&PRT&1400&157525.13&219308125506.737 \\
152&Qatar&QAT&421&68578.38&166928571428.571 \\
153&Reunion&REU&12&183.4&- \\
154&Romania&ROU&1333&85330.8500000001&211406933991.363 \\
155&Russia&RUS&3913&1340757.433&1578624060588.26 \\
156&Rwanda&RWA&34&261&9135454442.14013 \\
157&Saint Kitts \& Nevis&KNA&4&-&992007403.125926 \\
158&Saint Lucia&LCA&20&96.3&1810139888.88889 \\
159&Saint Vincent and The Grenadines&VCT&7&-&785222509.144563 \\
160&San Marino&SMR&5&-&1632860040.56795 \\
161&Saudi Arabia&SAU&1161&204125.646&688586133333.333 \\
162&Senegal&SEN&47&2468.5&21081669870.0624 \\
163&Serbia&SRB&360&11945.42&44120424391.86 \\
164&Seychelles&SYC&28&284.1&1503168689.81984 \\
165&Sierra Leone&SLE&23&-&3739577973.23943 \\
166&Singapore&SGP&6354&1782500.835&338406474038.67 \\
167&Slovakia&SVK&472&70136.38&95617670260.1145 \\
168&Slovenia&SVN&384&29887.97&48455919386.0505 \\
169&Solomon Islands&SLB&7&62.2&1321131090.73503 \\
170&Somalia&SOM&10&-&7128000000 \\
171&South Africa&ZAF&4417&833255.578&348871647962.321 \\
172&Spain&ESP&4030&1093061.456&1314314164402.2 \\
173&Sri Lanka&LKA&1502&29295.567&88019706803.834 \\
174&Sudan&SDN&60&970.9&123053386001.137 \\
175&Suriname&SUR&14&368.3&3068766109.75333 \\
176&Swaziland&SWZ&33&552.2&4433664364.24725 \\
177&Sweden&SWE&4978&894068.050000001&535607385506.432 \\
178&Switzerland&CHE&3604&4488854.00900001&678965423322.021 \\
179&Syria&SYR&68&1368.4&-\\\
180&Taiwan&TWN&4189&1168175.276&- \\
181&Tajikistan&TJK&24&-&7157865188.25222 \\
182&Tanzania&TZA&162&2866.22&53320625958.5628 \\
183&Thailand&THA&5445&651112.432999999&455275517239.347 \\
184&Togo&TGO&17&1200.9&4765866980.38429 \\
185&Tonga&TON&7&-&430174168.740104 \\
186&Trinidad and Tobago&TTO&127&9655.08&22250455018.8067 \\
187&Tunisia&TUN&149&6308.4&39952095560.8829 \\
188&Turkey&TUR&2488&404962.46&851549231502.615 \\
189&Turkmenistan&TKM&10&-&37926285714.2857 \\
190&Turks \& Caicos Islands&TCA&13&-&962525840 \\
191&Tuvalu&TUV&1&-&40620557.1335093 \\
192&Uganda&UGA&91&879.62&25995031850.1545 \\
193&Ukraine&UKR&734&47476.3630000001&112190355158.178 \\
194&United Arab Emirates&ARE&2121&226929.05&382575085091.899 \\
195&United Kingdom&GBR&24727&5795175.525&2637866340434.13 \\
196&United States of America&USA&121475&29655109.9389996&19485393853000 \\
197&Uruguay&URY&176&22385.08&56488991831.0239 \\
198&Uzbekistan&UZB&52&6061.7&59159949231.4924 \\
199&Vanuatu&VUT&11&9.46&849708342.698412 \\
200&Vatican City&VAT&3&-&- \\
201&Venezuela&VEN&262&195097.56&-\\\
202&Vietnam&VNM&2036&181079.663&223779865815.183 \\
203&Western Samoa&WSM&33&-&841538412.998896 \\
204&Yemen&YEM&52&609.664&26818703092.5852 \\
205&Zambia&ZMB&138&7017.01&25868142076.7897 \\
206&Zimbabwe&ZWE&162&4179.9&22813010116.1292 \\
207&Unknown&-&32907&3757.57&-\\
\hline
\caption{
{\tiny 
Countries and firm distribution. ``-" in some fields indicates missing information}} % needs to go inside longtable environment
\label{tab:country}
\end{longtable}
}
%-------------------------------------------------
%-------------------------------------------
\begin{longtable}{|r|l|r|}
\hline
Sl. no.&Name of the sector& No. of firms\\
\hline
1&Communication Services&17491\\
2&Consumer Discretionary&35283\\
3&Consumer Staples&13044\\
4&Energy&9542\\
5&Financials&25953\\
6&Health Care&20979\\
7&Industrials&54819\\
8&Information Technology&34648\\
9&Materials&18019\\
10&Real Estate&7555\\
11&Utilities&7017\\
12&Unknown&193103\\
\hline
\caption{
Sectors and firm distribution} % needs to go inside longtable environment
\label{tab:sectors}
\end{longtable}
%-------------------------------------------------

%--------------------------
\newpage
%-------------------------------------------------
%-------------------------------------------
\begin{longtable}{|r|l|l|}
\hline
ID&Primary industry&Sector \\
\hline
1&Advertising&Communication Services \\
2&Aerospace and Defense&Industrials \\
3&Agricultural Products&Consumer Staples \\
4&Agricultural and Farm Machinery&Industrials \\
5&Air Freight and Logistics&Industrials \\
6&Airlines&Industrials \\
7&Airport Services&Industrials \\
8&Alternative Carriers&Communication Services \\
9&Aluminum&Materials \\
10&Apparel Retail&Consumer Discretionary \\
11&Apparel, Accessories and Luxury Goods&Consumer Discretionary \\
12&Application Software&Information Technology \\
13&Asset Management and Custody Banks&Financials \\
14&Auto Parts and Equipment&Consumer Discretionary \\
15&Automobile Manufacturers&Consumer Discretionary \\
16&Automotive Retail&Consumer Discretionary \\
17&Biotechnology&Health Care \\
18&Brewers&Consumer Staples \\
19&Broadcasting&Communication Services \\
20&Building Products&Industrials \\
21&Cable and Satellite&Communication Services \\
22&Casinos and Gaming&Consumer Discretionary \\
23&Coal and Consumable Fuels&Energy \\
24&Commercial Printing&Industrials \\
25&Commodity Chemicals&Materials \\
26&Communications Equipment&Information Technology \\
27&Computer and Electronics Retail&Consumer Discretionary \\
28&Construction Machinery and Heavy Trucks&Industrials \\
29&Construction Materials&Materials \\
30&Construction and Engineering&Industrials \\
31&Consumer Electronics&Consumer Discretionary \\
32&Consumer Finance&Financials \\
33&Copper&Materials \\
34&Data Processing and Outsourced Services&Information Technology \\
35&Department Stores&Consumer Discretionary \\
36&Distillers and Vintners&Consumer Staples \\
37&Distributors&Consumer Discretionary \\
38&Diversified Banks&Financials \\
39&Diversified Capital Markets&Financials \\
40&Diversified Chemicals&Materials \\
41&Diversified Metals and Mining&Materials \\
42&Diversified REITs&Real Estate \\
43&Diversified Real Estate Activities&Real Estate \\
44&Diversified Support Services&Industrials \\
45&Drug Retail&Consumer Staples \\
46&Education Services&Consumer Discretionary \\
47&Electric Utilities&Utilities \\
48&Electrical Components and Equipment&Industrials \\
49&Electronic Components&Information Technology \\
50&Electronic Equipment and Instruments&Information Technology \\
51&Electronic Manufacturing Services&Information Technology \\
52&Environmental and Facilities Services&Industrials \\
53&Fertilizers and Agricultural Chemicals&Materials \\
54&Financial Exchanges and Data&Financials \\
55&Food Distributors&Consumer Staples \\
56&Food Retail&Consumer Staples \\
57&Footwear&Consumer Discretionary \\
58&Forest Products&Materials \\
59&Gas Utilities&Utilities \\
60&General Merchandise Stores&Consumer Discretionary \\
61&Gold&Materials \\
62&Health Care Distributors&Health Care \\
63&Health Care Equipment&Health Care \\
64&Health Care Facilities&Health Care \\
65&Health Care REITs&Real Estate \\
66&Health Care Services&Health Care \\
67&Health Care Supplies&Health Care \\
68&Health Care Technology&Health Care \\
69&Heavy Electrical Equipment&Industrials \\
70&Highways and Railtracks&Industrials \\
71&Home Furnishings&Consumer Discretionary \\
72&Home Improvement Retail&Consumer Discretionary \\
73&Homebuilding&Consumer Discretionary \\
74&Homefurnishing Retail&Consumer Discretionary \\
75&Hotel and Resort REITs&Real Estate \\
76&Hotels, Resorts and Cruise Lines&Consumer Discretionary \\
77&Household Appliances&Consumer Discretionary \\
78&Household Products&Consumer Staples \\
79&Housewares and Specialties&Consumer Discretionary \\
80&Human Resource and Employment Services&Industrials \\
81&Hypermarkets and Super Centers&Consumer Staples \\
82&IT Consulting and Other Services&Information Technology \\
83&Independent Power Producers and Energy Traders&Utilities \\
84&Industrial Conglomerates&Industrials \\
85&Industrial Gases&Materials \\
86&Industrial Machinery&Industrials \\
87&Industrial REITs&Real Estate \\
88&Insurance Brokers&Financials \\
89&Integrated Oil and Gas&Energy \\
90&Integrated Telecommunication Services&Communication Services \\
91&Interactive Home Entertainment&Communication Services \\
92&Interactive Media and Services&Communication Services \\
93&Internet Services and Infrastructure&Information Technology \\
94&Internet and Direct Marketing Retail&Consumer Discretionary \\
95&Investment Banking and Brokerage&Financials \\
96&Leisure Facilities&Consumer Discretionary \\
97&Leisure Products&Consumer Discretionary \\
98&Life Sciences Tools and Services&Health Care \\
99&Life and Health Insurance&Financials \\
100& Managed Health Care&Health Care \\
101& Marine&Industrials \\
102& Marine Ports and Services&Industrials \\
103& Metal and Glass Containers&Materials \\
104& Mortgage REITs&Financials \\
105&Motorcycle Manufacturers&Consumer Discretionary \\
106&Movies and Entertainment&Communication Services \\
107&Multi-Sector Holdings&Financials \\
108&Multi-Utilities&Utilities \\
109&Multi-line Insurance&Financials \\
110&Office REITs&Real Estate \\
111&Office Services and Supplies&Industrials \\
112&Oil and Gas Drilling&Energy \\
113&Oil and Gas Equipment and Services&Energy \\
114&Oil and Gas Exploration and Production&Energy \\
115&Oil and Gas Refining and Marketing&Energy \\
116&Oil and Gas Storage and Transportation&Energy \\
117&Other Diversified Financial Services&Financials \\
118&Packaged Foods and Meats&Consumer Staples \\
119&Paper Packaging&Materials \\
120&Paper Products&Materials \\
121&Personal Products&Consumer Staples \\
122&Pharmaceuticals&Health Care \\
123&Precious Metals and Minerals&Materials \\
124&Property and Casualty Insurance&Financials \\
125&Publishing&Communication Services \\
126&Railroads&Industrials \\
127&Real Estate Development&Real Estate \\
128&Real Estate Operating Companies&Real Estate \\
129& Real Estate Services&Real Estate \\
130& Regional Banks&Financials \\
131&Reinsurance&Financials \\
132& Renewable Electricity&Utilities \\
133&Research and Consulting Services&Industrials \\
134& Residential REITs&Real Estate \\
135& Restaurants&Consumer Discretionary \\
136& Retail REITs&Real Estate \\
137& Security and Alarm Services&Industrials \\
138&Semiconductor Equipment&Information Technology \\
139& Semiconductors&Information Technology \\
140& Silver&Materials \\
141& Soft Drinks&Consumer Staples \\
142& Specialized Consumer Services&Consumer Discretionary \\
143& Specialized Finance&Financials \\
144& Specialized REITs&Real Estate \\
145& Specialty Chemicals&Materials \\
146& Specialty Stores&Consumer Discretionary \\
147& Steel&Materials \\
148& Systems Software&Information Technology \\
149& Technology Distributors&Information Technology \\
150& Technology Hardware, Storage and Peripherals&Information Technology \\
151& Textiles&Consumer Discretionary \\
152& Thrifts and Mortgage Finance&Financials \\
153&Tires and Rubber&Consumer Discretionary \\
154& Tobacco&Consumer Staples \\
155& Trading Companies and Distributors&Industrials \\
156& Trucking&Industrials \\
157& Water Utilities&Utilities \\
158& Wireless Telecommunication Services&Communication Services \\
\hline
\caption{
Primary industry and sector classification} % needs to go inside longtable environment
\label{tab:primaryindustry}
\end{longtable}
%-------------------------------------------------

We show a different color code of the nodes for the overexpression network of primary industries, which is shown in  Fig~6 in main text. Here we use the node color according to their sector classification. From Fig~6 of main text and Fig~\ref{fig:sector}, we observe the clustering among primary industries are formed based their sectors.    
%--------------------------
\begin{figure}[!h]
\centering
\includegraphics[width=\linewidth]{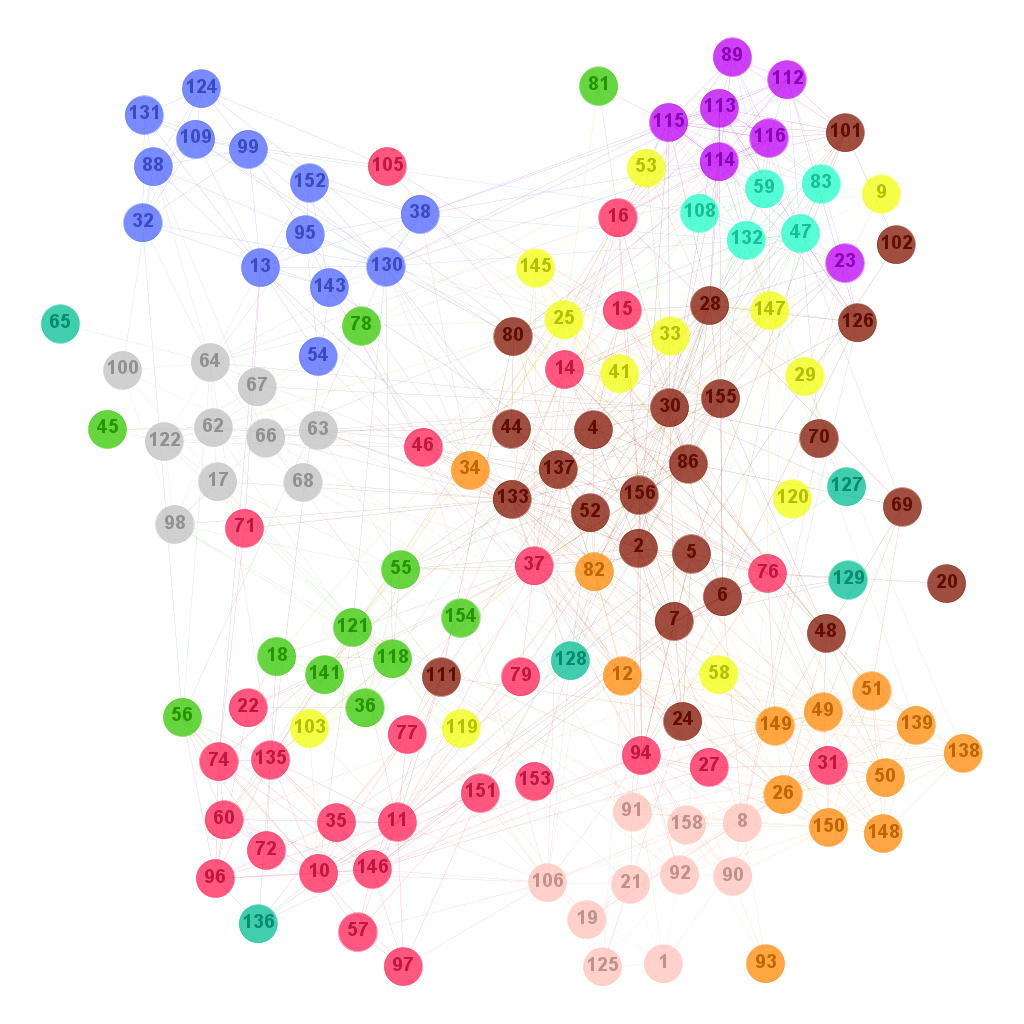}
\caption{{\bf  
Overexpression network of primary industries. Different node color indicates different sectors. IDs of the nodes are given in Table~\ref{tab:primaryindustry}.}
}
\label{fig:sector}
\end{figure}

%----------------------------------------------------------------------------------
%\input{9_supplement_1.tex}
%\clearpage
%\input{9_supplement_4.tex}
%\clearpage
%\input{9_supplement_2.tex}
%\clearpage
%\input{9_supplement_3.tex}
%\begin{thebibliography}{10}

%\end{thebibliography}

%\end{document}
%==================================================================================================%

\end{document}